\documentclass[journal]{IEEEtran}

\usepackage{graphicx}
\usepackage{cite}
\usepackage[cmex10]{amsmath}
\usepackage{amssymb}
\usepackage{mathtools}
\usepackage{siunitx}
\usepackage{subcaption}
\usepackage{xcolor}

\usepackage{booktabs}
\usepackage{tikz}
\usepackage{pgfplots}
\pgfplotsset{compat=newest}% <-- moves axis labels near ticklabels (respects tick label widths)
\pgfplotsset{legend style={rounded corners=2pt,nodes=right}}% <-- global figure style
\usetikzlibrary{arrows,decorations,decorations.markings,shapes}

\usepackage{balance}

\DeclareMathAlphabet{\mathbit}{OML}{cmr}{bx}{it}

\DeclareMathOperator{\e}{e}
\DeclareMathOperator{\E}{E}
\DeclareMathOperator{\T}{T}
\DeclareMathOperator{\Tr}{tr}

\DeclareMathOperator{\Probability}{Pr}

\DeclareMathOperator*{\argmin}{arg\,min}

\renewcommand\vec[1]{\operatorname{vec}\left(#1\right)}

\DeclareMathOperator{\fieldR}{\mathbb{R}}

\DeclareMathOperator{\fieldN}{\mathbb{N}}
\DeclareMathOperator{\fieldB}{\mathbb{B}}

\DeclareMathOperator{\distN}{\mathcal{N}}

\newcommand\vecT[1]{\operatorname{vec}^{\T}\left(#1\right)}

\newcommand{\sign}[1]{\Sign{\left(#1\right)}}

\newcommand{\ve}[1]{\boldsymbol{#1}}

\newcommand{\exdi}[2]{\E_{#1} \left[#2\right]}

\renewcommand{\exp}[1]{\operatorname{exp}\left(#1\right)}

\newcommand{\tr}[1]{\Tr \left(#1\right)}
\newcommand{\Prob}[1]{\Probability\left\{#1\right\}}

\newcommand\Sign{\operatorname{sign}}

\newcommand{\sinc}[1]{\operatorname{sinc} \left(#1\right)}

\title{Latency Analysis for Sequential Detection\\in Low-Complexity Binary Radio Systems}

\author{Manuel~S.~Stein and Michael~Fau{\ss}
\thanks{This work was in part funded by the Deutsche Forschungsgemeinschaft (DFG, German Research Foundation) - grant no. 413008418 and 424522268.}
\thanks{M. S. Stein is with the Department of Microelectronics, Technische Universiteit Delft, The Netherlands (e-mail: M.S.Stein@tudelft.nl). M. Fau{\ss} is with the Department of Electrical Engineering, Princeton University, USA (e-mail: mfauss@princeton.edu).}
}
\begin{document}
\maketitle
\begin{abstract}
We consider the problem of making a quick decision in favor of one of two possible physical signal models while the numerical measurements are acquired by sensing devices featuring minimal digitization complexity. Therefore, the digital data streams available for statistical processing are binary and exhibit temporal and spatial dependencies. To handle the intractable multivariate binary data model, we first consider sequential tests for exponential family distributions. Within this generic probabilistic framework, we identify adaptive approximations for the log-likelihood ratio and the Kullback-Leibler divergence. The results allow designing sequential detectors for binary radio systems and analyzing their average run-time along classical arguments of Wald. In particular, the derived tests exploit the spatio-temporal correlation structure of the analog sensor signals engraved into the binary measurements. As an application, we consider the specification of binary sensing architectures for cognitive radio and GNSS spectrum monitoring where our results characterize the sequential detection latency as a function of the temporal oversampling and the number of antennas. Finally, we evaluate the efficiency of the proposed algorithms and illustrate the accuracy of our analysis via Monte-Carlo simulations.
\end{abstract} 
\section{Introduction}\label{sec:introduction}
The design of future sensor systems represents a challenge. For applications in the Internet of Things (IoT), the focus is on further miniaturization. Thus, small circuit size, low production cost, and low energy consumption are essential requirements. In contrast, for safety-critical applications, sensing accuracy and detection reliability are of utmost importance. In any case, optimal system design either aims at achieving minimum complexity at a specified performance level, or at converting the available resources into maximum performance. Advancing sensor technology, therefore, requires a thorough understanding of Pareto-optimal architectures, i.e., sensing systems for which it is impossible to improve on complexity or performance without sacrificing the other measure.

A difficulty is that terms like complexity and performance are fuzzy without restriction to a specific perspective. In an R\&D environment, analog front-end engineers tend to equate complexity with the circuit area or the dissipated power, and performance with the degree of linearity. On the other hand, software engineers instead associate complexity with the computing effort and the size of the occupied memory, while performance for them is linked to the fast and correct response of the digital units to specific input data. In hardware-aware statistical signal processing, the understanding is emerging that a holistic approach is required when designing advanced sensor systems. In particular, this means that the physical phenomenon, the design of the analog front-end, and digital processing should be considered as a single joint problem.

For this purpose, it is helpful to reduce the system task to its fundamental building blocks. Elementary for sensing are parameter estimation and signal detection. While estimation aims at inferring data model parameters within an open set from noisy measurements, detection covers cases with discrete parameter space. To characterize performance, the accuracy in determining the parameters is central within estimation. Detection theory focuses on the reliability in discriminating between the possible data models. An advantage of the parametric interpretation within these disciplines is that there is an understanding of optimal procedures. Also, analytic expressions are available to characterize the achievable accuracy or reliability. Hence, for optimal system design, it is desirable to formulate a suitable probabilistic measurement data model as a function of the acquisition apparatus and the physical effects acting on it. Together with technology cost models, physical sensor layouts can be determined, which offer a favorable complexity-performance trade-off. Based on these considerations, system engineers can give detailed design recommendations to analog front-end engineers and digital software engineers. 
\subsection{Motivation}
The purpose of this article is to highlight the opportunities and challenges of a hardware-aware system design approach. We limit ourselves to sensing systems where the analog-to-digital (A/D) converters only forward the information concerning the sign of the analog sensor amplitude to the digital processing units. The reasoning is that amplitude resolution increases A/D complexity exponentially. Each additional bit at least doubles the demand on A/D resources \cite{MurmannSurvey}. Thus, from a complexity-aware perspective, digitization with more than $1$-bit amplitude resolution appears unfavorable, in particular, for applications where most bits represent uninformative noise. Furthermore, binary digitization is beneficial for low-level processing, where data can be handled with low logic complexity. However, these savings are obtained by accepting a considerable loss of information during signal acquisition. Fortunately, a thorough analysis shows that probabilistic modeling of the transition from a physical phenomenon to binary measurements, front-end optimization, and likelihood-oriented data processing can compensate for the effect of digitization-induced distortion on the final inference solution, see, e.g., \cite{SteinICASSP17}. While the potential of binary sampling has been extensively studied, e.g., regarding wireless communication capabilities \cite{Ivrlac06,Dabeer06,Mezghani07,Mezghani12,Mo15,Jacobsson15,Mollen17,Jeon18}, signal reconstruction error \cite{Daubechies03,Kamilov12,Boufounos15}, estimation sensitivity \cite{Madsen00,Ribeiro06_part1,Mezghani10,Stein15_WCL,Li17,Mezghani18}, and detection reliability \cite{Willett95,Ciuonzo13,Stein18_ICASSP}, here we focus on the analysis of the sensing latency \cite{Tantaratana77}.

To this end, the binary data stream is assumed to be processed in short spatio-temporal blocks to quickly detect which of two possible physical scenarios is generating the measurements. The reliability level to be achieved is predetermined. Leaning on concepts from sequential analysis \cite{Wald45, Fauss15}, we characterize the average number of blocks required for reliable detection when binary sensors are employed. A challenge that we address is the fact that the probabilistic data models characterizing multivariate binary measurements are, in general, intractable. A workaround, aiming at approximating likelihood-based sequential testing, enables analyzing the detection latency as a function of the binary sensing architecture and the two underlying physical models. 

Applications related to mobile communication and satellite-based synchronization systems are outlined. On the one hand, we consider cognitive radio where a secondary user observes the activity of a primary user \cite{Axell12}. Once the primary wireless transmitter is inactive, the secondary system uses the communication channel. Thus, without limiting the functionality of the primary user or occupying additional bands, spectral resources can be used for wireless services. Since mobile radio systems necessitate a cost-efficient and miniaturized design, reliable operation at low complexity is crucial. On the other hand, we treat the monitoring of safety-critical global navigation satellite system (GNSS) frequencies \cite{Broumandan16}. Due to the low power and the distance of the GNSS transmitters, terrestrial radio interference poses a challenge. Small jamming devices, which protect the privacy of individuals, can impair the functionality of critical infrastructure (e.g., financial markets, power networks, airports, communication networks), which ensures the safety and well-being of many people \cite{Ioannides16}. For monitoring the GNSS spectrum in the vicinity of safety-critical receivers, high performance is of utmost importance. 

Both applications have in common that during system development, a favorable complexity-performance trade-off has to be identified. In addition to reaching reliability, minimizing latency plays a decisive role. A cognitive system with a small decision delay enables using spectral resources efficiently. Monitoring sensors that quickly detect interference can initiate measures for suppression or declare malfunctioning in time. For such architectures, we perform hardware-aware modeling of the analog front-ends and the resulting binary measurements. Our findings characterize how the temporal oversampling and the number of sensors affect the sequential detection latency. Additionally, we use A/D cost models to identify cases in which it might be advantageous to deploy higher A/D resolution. We close the discourse by evaluating the developed sequential algorithms and illustrating the accuracy of our latency analysis utilizing synthetically generated data streams from exemplary binary radio systems.

\subsection{Related Work}
Quantized sequential decision-making has been studied predominantly for sensor networks where measurement nodes forward compressed statistics of their observations to minimize the communication overhead \cite{Hashemi89, Veeravalli93, Hussain94, Mei08, Yilmaz12, Chaudhari12, Wang13}. In such a setup, it is usually assumed that the sensing nodes have access to the unquantized digital observations and compress them, for example, by quantizing the likelihood ratio. Here we discuss sequential detection when employing binary sensing front-ends. Therefore, the nodes do not have access to the high-resolution observations, and the centralized decision is exclusively based on the hard-limited measurements of all sensing nodes. This case is less well studied, and the focus of existing works is on the design of optimal quantizers \cite{Tantaratana77,Blum95,Nguyen08,Teng13}. Note that works studying the sensing latency in centralized sequential detection with quantized measurements usually consider data models with deterministic signals in white noise, see, e.g., \cite{Tantaratana77,Tantaratana77_TIT, Lee81}.
\subsection{Contribution}
In contrast, we consider quantizers of minimal complexity. Therefore, the analog signals are converted into binary data by symmetric hard-limiting. Furthermore, we assume random signals with temporal and spatial dependencies. Sequential analysis then exhibits particular challenges which, to the best of our knowledge, have not been addressed yet. Due to the highly nonlinear signal acquisition, for rigorous analysis, one cannot rely on linear system theory and Gaussian statistics when modeling the sensor data. Such methods only provide accurate results when a sufficiently high A/D resolution is deployed. On the other hand, with likelihood-based approaches, see, e.g., \cite{Choi16,Hong18,SteinBar18}, the analysis becomes challenging, when considering correlation at the hard-limiter input. For the resulting multivariate binary data, the mass function and the sufficient statistics, in general, grow exponentially with the number of variables \cite{Dai13}. Furthermore, the unknown orthant probabilities \cite{Gupta63} with more than four variables (an open problem in statistics) hinder access to the likelihood. We tackle this by exploiting the properties of the exponential family. Such a generic probabilistic perspective onto hardware-aware signal processing systems \cite{Stein_FisherBound} provides an approximate log-likelihood ratio (ALLR) and, therefore, enables performing a sequential probability ratio test (SPRT) for a broad class of data models without direct access to the likelihood ratio. Additionally, the approach provides tuneable approximations for the Kullback-Leibler divergence characterizing the average sampling number (ASN) of the approximate SPRT (ASPRT). Further, using an auxiliary distribution of reduced statistical complexity, we ensure computational tractability. By studying the efficiency of binary radio sensor layouts via our results, we show that with temporal and spatial oversampling, these systems master challenging sequential detection tasks. In particular, the possibility to deploy more antennas enables sensing with a superior complexity-performance trade-off. Note that this article is an extension of our special session contribution \cite{SteinFauss18}, where we discussed preliminary results.
\section{Problem Formulation}\label{sec:problem:form}
\subsection{System Model - Signal Acquisition}
We consider $M\in\fieldN$ analog sensor outputs, modeled as real-valued time-continuous functions
\begin{align}\label{receive:signal:analog}
\ve{y}(t)=\ve{s}(t)+\ve{\eta}(t).
\end{align}
The analog measurements $\ve{y}(t)\in\fieldR^M, t\in\fieldR,$ are the superposition of a source component $\ve{s}(t)\in\fieldR^M$ and additive independent measurement noise $\ve{\eta}(t)\in\fieldR^M$. Both signals are modeled as band-limited wide-sense stationary Gaussian processes with zero mean. The analog signals \eqref{receive:signal:analog} are synchronously discretized in batches of $K\in\fieldN$ equidistant sampling points in time. With infinite digital amplitude resolution, the $n$th observation instance provides a space-time sample
\begin{align}\label{def:space:time:sample}
\ve{y}_n=\begin{bmatrix} \ve{y}^{\T}_n[1] &\ve{y}^{\T}_n[2] &\ldots &\ve{y}^{\T}_n[K] \end{bmatrix}^{\T} \in \fieldR^{MK},
\end{align}
where, with the sampling interval $T\in\fieldR$, we write
\begin{align}
\ve{y}_n[k]=\ve{y}\big((n-1)KT+(k-1)T\big), \quad k=1,\ldots,K.
\end{align}
Note, that the sampling duration for each block \eqref{def:space:time:sample} is $T_o=KT$. We consider the dependencies between consecutive samples as negligible, such that, for each $n\in\fieldN$, \eqref{def:space:time:sample} can be considered to be independent. Due to stationarity and Gaussianity, the spatio-temporal data \eqref{def:space:time:sample} follows the model
\begin{align}\label{digital:model:multivariate:gauss}
\ve{y}=\ve{s}+\ve{\eta},\quad\ve{y},\ve{s},\ve{\eta}\in\fieldR^{MK},\quad\ve{y}\sim \distN(\ve{0}, \ve{R}_{\ve{y}}(\ve{\theta})),
\end{align}
where the space-time covariance matrix
\begin{align}
\ve{R}_{\ve{y}}(\ve{\theta})&=\exdi{\ve{y};\ve{\theta}}{\ve{y}\ve{y}^{\T}}\notag\\
&=\ve{R}_{\ve{s}}(\ve{\theta})+\ve{R}_{\ve{\eta}}, \quad \ve{R}_{\ve{y}}(\ve{\theta})\in\fieldR^{MK \times MK}
\end{align}
is a superposition of the source covariance $\ve{R}_{\ve{s}}(\ve{\theta})=\exdi{\ve{s};\ve{\theta}}{\ve{s}\ve{s}^{\T}}$ and the noise covariance $\ve{R}_{\ve{\eta}}=\exdi{\ve{\eta}}{\ve{\eta}\ve{\eta}^{\T}}$. Note that the source covariance is a function of the parameters $\ve{\theta}\in\fieldR^D$, while the noise covariance is constant.

Realizing a signal acquisition, which approximately produces a data stream according to the multivariate Gaussian model \eqref{digital:model:multivariate:gauss}, in practice, requires an A/D converter with several bits digital amplitude resolution at each analog output. To minimize A/D complexity, we here assume that within the considered system, only the signs of the analog measurements \eqref{receive:signal:analog} are transferred to the digital processing units. Such a binary signal acquisition results in the space-time observations
\begin{align}\label{system:model:sign}
\ve{z}_n = \sign{\ve{y}_n},
\end{align}
where the element-wise hard-limiter $\sign{\cdot}$ is defined
\begin{align}
\left[\ve{z}_n\right]_i=
\begin{cases}
+1& \text{if } [\ve{y}_n]_i \geq 0,\\
-1 & \text{if } [\ve{y}_n]_i < 0.
\end{cases}
\end{align}
Per space-time sample \eqref{def:space:time:sample}, analog-to-binary (A/B) conversion \eqref{system:model:sign} can be realized by $K$ comparator operations for each of the $M$ analog outputs, while a $b$-bit receiver requires to activate 
\begin{align}\label{def:sample:cost}
\text{SC}^{(b)}(M,K)=MK(2^b-1),\quad b\in\fieldN,
\end{align}
comparators. Further, the binary data \eqref{system:model:sign} can be stored on small memory, transmitted using channels with moderate capacity, and preprocessed at a high rate and low computational cost.  
\subsection{Processing Task - Sequential Decision-Making}
The binary measurements \eqref{system:model:sign} gathered up to the $n$th observation instance, are summarized
\begin{align}\label{quantized:data}
\ve{Z}_n=\begin{bmatrix} \ve{z}_1 &\ve{z}_2 &\ldots &\ve{z}_n\end{bmatrix}, \quad\ve{Z}_n\in\fieldB^{MK\times n}.
\end{align}
The inference task is to use the available data $\ve{Z}_n$ to decide which of the two possible probability laws
\begin{equation}\label{receive:hypotheses}
\mathcal{H}_0\colon \ve{z}\sim p_{\ve{z}}(\ve{z};\ve{\theta}_0)
\quad \text{or} \quad
\mathcal{H}_1\colon \ve{z}\sim p_{\ve{z}}(\ve{z};\ve{\theta}_1)
\end{equation}
is the model generating the output data \eqref{system:model:sign}. The detection is to be conducted reliably, i.e.,
\begin{align}\label{prob:errors}
&\Prob{\text{decision }\mathcal{H}_0 | \mathcal{H}_1 }\leq \alpha_0,\\
&\Prob{\text{decision }\mathcal{H}_1 | \mathcal{H}_0 }\leq \alpha_1.
\end{align}
If a decision based on $\ve{Z}_n$ would lead to a violation of these reliability constraints, the processing unit waits for the next space-time sample $\ve{z}_{n+1}$ and tries to perform the test with the augmented data stream $\ve{Z}_{n+1}$. The instance in which the detection is finally performed is denoted by $n_{\text{D}}\in\fieldN$. The performance of the sequential test is characterized by the average sampling number (ASN), which is defined as the expected value of $n_{\text{D}}$ under the data-generating model $p_{\ve{z}}(\ve{z};\ve{\theta})$, i.e.,
\begin{align}\label{def:asn}
\text{ASN}(\ve{\theta})=\exdi{n_{\text{D}};\ve{\theta}}{n_{\text{D}}}.
\end{align}
Note that the ASN characterizes expected latency, i.e., individual decisions might require less or more samples \eqref{system:model:sign}.
A classical approach to construct a decision-making algorithm minimizing \eqref{def:asn} is the sequential probability ratio test (SPRT) \cite{Wald45}. Given the data stream $\ve{Z}_n$, the log-likelihood ratio (LLR)
\begin{align}\label{def:LLR:data}
l(\ve{Z}_n) = \sum_{i=1}^n l(\ve{z}_i) = \sum_{i=1}^n \ln{ \frac{p_{\ve{z}}(\ve{z}_i;\ve{\theta}_1)}{p_{\ve{z}}(\ve{z}_i;\ve{\theta}_0)} }
\end{align}
is calculated and compared against two decision thresholds. If
\begin{align}\label{sprt:thresh:low}
l(\ve{Z}_n)\leq\ln{\frac{\alpha_1}{1-\alpha_0}}=L_0,
\end{align}
the test is stopped with a decision in favor of $\mathcal{H}_0$. In case 
\begin{align}\label{sprt:thresh:up}
l(\ve{Z}_n)\geq \ln{\frac{1-\alpha_1}{\alpha_0}}=L_1,
\end{align}
the sequential decision-making is terminated in favor of the hypothesis $\mathcal{H}_1$. Otherwise, an additional signal sample $\ve{z}_{n+1}$ is taken to continue the test with $\ve{Z}_{n+1}$. With the short notations
\begin{align}
N_0&=(1-\alpha_0) \ln{\frac{\alpha_1}{1-\alpha_0}} + \alpha_0\ln{\frac{1-\alpha_1}{\alpha_0}},\\
N_1&=\alpha_1 \ln{\frac{\alpha_1}{1-\alpha_0}} + \big(1-\alpha_1\big)\ln{\frac{1-\alpha_1}{\alpha_0}},
\end{align}
the ASN of the SPRT under the two possible data model hypotheses \eqref{receive:hypotheses} is approximately \cite{Wald45}
\begin{align}
\label{ASN0}
\text{ASN}_0 &\approx \frac{ N_0}{ \exdi{\ve{z};\ve{\theta}_0}{ l(\ve{z}) }} =-\frac{ N_0}{ D(p_{\ve{z};\ve{\theta}_0}||p_{\ve{z};\ve{\theta}_1}) }
\end{align}
and
\begin{align}
\label{ASN1}
\text{ASN}_1 \approx \frac{N_1}{ \exdi{\ve{z};\ve{\theta}_1}{ l(\ve{z}) } }=\frac{N_1}{ D(p_{\ve{z};\ve{\theta}_1}||p_{\ve{z};\ve{\theta}_0}) }, 
\end{align}
where $D(p_{\ve{u};\ve{\theta}}||q_{\ve{w};\ve{\theta}})$ denotes the Kullback--Leibler divergence between the distributions $p_{\ve{u}}(\ve{u};\ve{\theta})$ and $q_{\ve{w}}(\ve{w};\ve{\theta})$.
\subsection{Challenge - Data Models for Binary Measurements}
While A/B conversion, according to \eqref{system:model:sign}, offers significant savings regarding hardware cost and power consumption, the probabilistic characterization of the binary sensor outputs forms a challenge. To obtain the exact binary likelihood ratio required in \eqref{def:LLR:data}, the multidimensional integral
\begin{align}\label{likelihood:quantizer}
p_{\ve{z}}(\ve{z};\ve{\theta})=\int_{\ve{\mathcal{Y}}(\ve{z})} p_{\ve{y}}(\ve{y};\ve{\theta}) {\mathrm d}\ve{y}
\end{align}
needs to be evaluated, where $p_{\ve{y}}(\ve{y};\ve{\theta})$ denotes the distribution of the data at the input to \eqref{system:model:sign} and
\begin{align}
\ve{\mathcal{Y}}(\ve{z}) = \left\{ \ve{y} \in \fieldR^{MK} \,\big|\, \ve{z} = \sign{\ve{y}} \right\}.
\end{align} 
The calculation of an integral like \eqref{likelihood:quantizer} can turn out to be challenging. If, like in our case, the input to \eqref{system:model:sign} is multivariate Gaussian, evaluation of \eqref{likelihood:quantizer} requires the orthant probabilities, for which solutions are only known for $MK\leq4$. Even if these probabilities were available, the memory required to store all possible values for a single hypothesis scales as $\mathcal{O}(2^{MK})$. This renders using \eqref{def:LLR:data} for sequential detection and analyzing the resulting ASN by \eqref{ASN0} and \eqref{ASN1} prohibitively complicated even for scenarios with moderately large $M$ and $K$.
\section{Likelihood Ratio in the Exponential Family}
A conceptual observation that turns out useful is that distributions of multivariate binary measurements \eqref{system:model:sign} can be represented within the framework of the exponential family.
\subsection{Exponential Family Data Models}
A distribution belongs to the exponential family if
\begin{align}\label{def:exp:family}
{p}_{\ve{u}}(\ve{u};\ve{\theta})=\exp{\ve{\beta}^{\T}(\ve{\theta}) \ve{\phi}(\ve{u}) - \lambda(\ve{\theta})+\nu(\ve{u})},
\end{align}
where $\ve{u}\in\ve{\mathcal{U}}$ is the $V$-variate data with support $\ve{\mathcal{U}}$, $\ve{\theta}\in\fieldR^D$ the physical parameters, $\ve{\beta}(\ve{\theta})\colon \fieldR^{D} \to \fieldR^{C}$ the statistical weights\footnote{Usually the term ``natural parameters'' is used for $\ve{\beta}(\ve{\theta})$. We use ``statistical weights'' to emphasize that, in our engineering-oriented perspective, a probabilistic data model ${p}_{\ve{u}}(\ve{u};\ve{\theta})$ forms the connection between an analog physical/natural phenomenon $\ve{\theta}$ and digital measurement data $\ve{u}$.}, $\ve{\phi}(\ve{u})\colon \ve{\mathcal{U}} \to\fieldR^{C}$ the sufficient statistics, $\lambda(\ve{\theta})\colon \fieldR^D \to \fieldR$ the log-normalizer and $\nu(\ve{u})\colon \ve{\mathcal{U}} \to\fieldR$ the carrier measure. While the multivariate Gaussian model \eqref{digital:model:multivariate:gauss} also factorizes according to \eqref{def:exp:family}, the number of its sufficient statistics $C$ scales as $\mathcal{O}(V^2)$. For multivariate binary data, in contrast, these statistics scale as $\mathcal{O}(2^V)$. This is due to the fact that in multivariate binary distributions the sufficient statistics not only comprise the pairwise products between all variables but also all higher-order products \cite{Dai13}. Therefore, the LLR
\begin{align}\label{def:LLR}
l(\ve{u})=\ln{ \frac{p_{\ve{u}}(\ve{u};\ve{\theta}_1)}{p_{\ve{u}}(\ve{u};\ve{\theta}_0)} },
\end{align}
required in \eqref{def:LLR:data}, can be inconvenient to handle. In the following, we discuss LLR approximations 
\begin{align}\label{definiton:llr:approximations}
\tilde{l}(\ve{u})\approx l(\ve{u}), 
\end{align}
enabling to implement the test defined in \eqref{def:LLR:data}--\eqref{sprt:thresh:up} and assess its latency \eqref{ASN0} and \eqref{ASN1} via
\begin{align}\label{concept:approx:ellr}
\exdi{\ve{u};\ve{\theta}_i}{ l(\ve{u}) } \approx \exdi{\ve{u};\ve{\theta}_i}{ \tilde{l}(\ve{u}) }, \quad i=0,1.
\end{align}

To this end, note that the exact LLR between two hypotheses within the exponential family \eqref{def:exp:family} is given by
\begin{align}\label{def:llr:exp:fam}
l(\ve{u}) =\big(\ve{\beta}(\ve{\theta}_1) - \ve{\beta}(\ve{\theta}_0)\big)^{\T} \ve{\phi}(\ve{u}) - \big(\lambda(\ve{\theta}_1) - \lambda(\ve{\theta}_0)\big),
\end{align}
such that, with the mean of the sufficient statistics
\begin{align}\label{exp:family:mean}
\ve{\mu}_{\ve{\phi}}(\ve{\theta}) = \exdi{\ve{u};\ve{\theta}}{ \ve{\phi}(\ve{u}) }
\end{align}
and their covariance matrix
\begin{align}\label{exp:family:covariance}
\ve{R}_{\ve{\phi}}(\ve{\theta})&=\exdi{\ve{u};\ve{\theta}}{ \big( \ve{\phi}(\ve{u}) - \ve{\mu}_{\ve{\phi}}(\ve{\theta}) \big) \big(\ve{\phi}(\ve{u})-\ve{\mu}_{\ve{\phi}}(\ve{\theta}) \big)^{\T}},
\end{align}
the mean of the LLR is given by
\begin{align}\label{llr:exp:fam:mean}
\mu_i&=\exdi{\ve{u};\ve{\theta}_i}{ l(\ve{u}) }\notag\\
&=\big(\ve{\beta}(\ve{\theta}_1) - \ve{\beta}(\ve{\theta}_0)\big)^{\T} \ve{\mu}_{\ve{\phi}}(\ve{\theta}_i)  - \big(\lambda(\ve{\theta}_1) - \lambda(\ve{\theta}_0)\big),
\end{align}
while the variance of \eqref{def:llr:exp:fam} can be written as
\begin{align}\label{llr:exp:fam:variance}
\sigma_i^2&={\exdi{\ve{u};\ve{\theta}_i}{ \big( l(\ve{u}) - \exdi{\ve{u};\ve{\theta}_i}{ l(\ve{u}) } \big)^2 }}\notag\\
&={\big(\ve{\beta}(\ve{\theta}_1) - \ve{\beta}(\ve{\theta}_0)\big)^{\T} \ve{R}_{\ve{\phi}}(\ve{\theta}_i) \big(\ve{\beta}(\ve{\theta}_1) - \ve{\beta}(\ve{\theta}_0)\big)}.
\end{align}
\subsection{Approximations for the Exponential Family LLR}
In practice, access to the statistical weights $\ve{\beta}(\ve{\theta})$ and the log-normalizer $\lambda(\ve{\theta})$, for executing and analyzing likelihood-based tests, can be challenging to obtain. For example, for the binary output \eqref{system:model:sign}, the integral providing the log-normalizer
\begin{align}\label{def:log:norm}
\lambda(\ve{\theta})= \ln \int_{\ve{\mathcal{U}}}\exp{\ve{\beta}^{\T}(\ve{\theta}) \ve{\phi}(\ve{u}) + \nu(\ve{u})} {\rm d}\ve{u}
\end{align}
results in a sum with $2^{V}$ terms. To obtain a representation of \eqref{def:llr:exp:fam} which does not require explicit access to $\ve{\beta}(\ve{\theta})$ and $\lambda(\ve{\theta})$, we assume to have at hand \eqref{exp:family:mean} and \eqref{exp:family:covariance} as functions of $\ve{\theta}$. These measures are usually easier to obtain than $\ve{\beta}(\ve{\theta})$ and $\lambda(\ve{\theta})$. To link \eqref{exp:family:mean} and \eqref{exp:family:covariance} to the exponential family LLR \eqref{def:llr:exp:fam}, we use that all distributions \eqref{def:exp:family} exhibit regularity, i.e.,
\begin{align}\label{def:regularity}
\exdi{\ve{u};\ve{\theta}}{ \frac{ \partial \ln{ p_{\ve{u}}(\ve{u};\ve{\theta})} }{ \partial \ve{\theta} }}=\ve{0}^{\T}.
\end{align}
Consequently, for any exponential family \eqref{def:exp:family}, it holds that
\begin{align}\label{const:regularity}
\bigg( \frac{\partial \lambda(\ve{\theta})}{\partial \ve{\theta}} \bigg)^{\T} = \bigg( \frac{\partial \ve{\beta}(\ve{\theta})}{\partial \ve{\theta}} \bigg)^{\T} \ve{\mu}_{\ve{\phi}}(\ve{\theta}).
\end{align}
Defining an LLR linearization point
\begin{align}\label{definition:LLR:lin:point}
\ve{\tilde{\theta}}(\xi) = \xi \ve{\theta}_0 + (1-\xi) \ve{\theta}_1,\quad \xi\in [0, 1],
\end{align}
and applying the finite difference approximation \eqref{dif:approx:flex:vec} in the Appendix together with the regularity constraint \eqref{const:regularity}, gives
\begin{align}\label{lambda:approx:step1}
\lambda(\ve{\theta}_1) - \lambda(\ve{\theta}_0) &\approx \frac{\partial \lambda \big( \ve{\tilde{\theta}}(\xi) \big) }{\partial \ve{\theta}} (\ve{\theta}_1-\ve{\theta}_0)\notag\\
&=\ve{\mu}^{\T}_{\ve{\phi}}(  \ve{\tilde{\theta}}(\xi) ) \frac{\partial \ve{\beta}\big( \ve{\tilde{\theta}}(\xi) \big)}{\partial \ve{\theta}}(\ve{\theta}_1-\ve{\theta}_0).
\end{align}
Further, for exponential distributions \eqref{def:exp:family}, one obtains \cite{Stein_FisherBound}
\begin{align}
\frac{\partial \ve{\beta}(\ve{\theta})}{\partial \ve{\theta}}=\ve{R}_{\ve{\phi}}^{-1}(\ve{\theta}) \frac{\partial \ve{\mu}_{\ve{\phi}}(\ve{\theta}) }{\partial \ve{\theta}},
\end{align}
such that \eqref{lambda:approx:step1} can be reformulated as
\begin{align}\label{lambda:approx}
\lambda(\ve{\theta}_1) - \lambda(\ve{\theta}_0) &\approx \ve{\mu}^{\T}_{\ve{\phi}}\big(  \ve{\tilde{\theta}}(\xi) \big) \ve{R}_{\ve{\phi}}^{-1}\big( \ve{\tilde{\theta}}(\xi) \big) \frac{\partial \ve{\mu}_{\ve{\phi}}\big( \ve{\tilde{\theta}}(\xi) \big) }{\partial \ve{\theta}} (\ve{\theta}_1-\ve{\theta}_0).
\end{align}
Accordingly, the difference of the statistical weights in \eqref{def:llr:exp:fam} can be approximated by
\begin{align}\label{beta:approx}
\ve{\beta}(\ve{\theta}_1) - \ve{\beta}(\ve{\theta}_0) &\approx \frac{\partial \ve{\beta} \big( \ve{\tilde{\theta}}(\xi) \big) }{\partial \ve{\theta}} (\ve{\theta}_1-\ve{\theta}_0)\notag\\
&=\ve{R}_{\ve{\phi}}^{-1}\big( \ve{\tilde{\theta}}(\xi) \big) \frac{\partial \ve{\mu}_{\ve{\phi}}\big( \ve{\tilde{\theta}}(\xi) \big) }{\partial \ve{\theta}} (\ve{\theta}_1-\ve{\theta}_0).
\end{align}
Defining an LLR hyperplane
\begin{align}\label{LLR:hyperplane:der}
\ve{b}_{\partial}(\ve{\theta}_0,\ve{\theta}_1;\xi)=\ve{R}_{\ve{\phi}}^{-1}\big( \ve{\tilde{\theta}}(\xi) \big) \frac{\partial \ve{\mu}_{\ve{\phi}}\big( \ve{\tilde{\theta}}(\xi) \big) }{\partial \ve{\theta}} (\ve{\theta}_1-\ve{\theta}_0),
\end{align}
and using \eqref{lambda:approx} and \eqref{beta:approx} in \eqref{def:llr:exp:fam}, provides the approximation
\begin{align}\label{lrt:approx:derivative}
l(\ve{u})&\approx\ve{b}^{\T}_{\partial}(\ve{\theta}_0,\ve{\theta}_1;\xi) \Big( \ve{\phi}(\ve{u}) - \ve{\mu}_{\ve{\phi}}\big( \ve{\tilde{\theta}}(\xi) \big) \Big)
=\tilde{l}_{\partial}(\ve{u};\xi).
\end{align}
The structure of this approximate LLR (ALLR) enables interpreting the LLR of exponential family distributions \eqref{def:llr:exp:fam} as the signed distance of a sufficent statistics residual from the hyperplane \eqref{LLR:hyperplane:der}. Note, that \eqref{LLR:hyperplane:der} requires access to the derivative of the mean \eqref{exp:family:mean} with respect to $\ve{\theta}$ evaluated at $\ve{\tilde{\theta}}(\xi)$. By using the finite difference \eqref{dif:approx:flex:vec} to eliminate the derivative in \eqref{LLR:hyperplane:der}, the LLR hyperplane can also be written as
\begin{align}\label{LLR:hyperplane}
\ve{b}(\ve{\theta}_0,\ve{\theta}_1;\xi)=\ve{R}_{\ve{\phi}}^{-1}\big( \ve{\tilde{\theta}}(\xi) \big)\big(\ve{\mu}_{\ve{\phi}}(\ve{\theta}_1)-\ve{\mu}_{\ve{\phi}}(\ve{\theta}_0)\big).
\end{align}
Therefore, an alternative to the LLR approximation \eqref{lrt:approx:derivative} is
\begin{align}\label{lrt:approx}
l(\ve{u})&\approx \ve{b}^{\T}(\ve{\theta}_0,\ve{\theta}_1;\xi) \Big( \ve{\phi}(\ve{u}) - \ve{\mu}_{\ve{\phi}}\big( \ve{\tilde{\theta}}(\xi) \big) \Big)
=\tilde{l}(\ve{u};\xi).
\end{align}
Evaluating the ALLR \eqref{lrt:approx} requires access to the mean \eqref{exp:family:mean} with respect to $p_{\ve{u}}(\ve{u};\ve{\theta}_0)$, $p_{\ve{u}}(\ve{u};\ve{\theta}_1)$, and $p_{\ve{u}}(\ve{u};\ve{\tilde{\theta}}(\xi))$. Further, one requires the covariance \eqref{exp:family:covariance} with respect to $p_{\ve{u}}(\ve{u};\ve{\tilde{\theta}}(\xi))$. Note that integrals like \eqref{likelihood:quantizer} and \eqref{def:log:norm} are not required for evaluating the ALLRs \eqref{lrt:approx:derivative} and \eqref{lrt:approx}.
\section{Approximate Tests in the Exponential Family}\label{sec:one:bit:sprt}
\subsection{Approximate Sequential Probability Ratio Test}
Defining the empirical mean of the sufficent statistics
\begin{align}\label{empirical:mean:suff:stat}
\ve{\hat{\mu}_{\ve{\phi}}}(\ve{U}_n)=\frac{1}{n} \sum_{i=1}^{n} \ve{\phi}(\ve{u}_i),
\end{align}
and using the LLR approximation \eqref{lrt:approx:derivative} or \eqref{lrt:approx}, for any data stream associated with an exponential family model \eqref{def:exp:family}, an approximate SPRT (ASPRT) can be performed by comparing
\begin{align}\label{lrr:exp:approx:der}
\tilde{l}_{\partial}(\ve{U}_n;\xi) &= \sum_{i=1}^{n}\tilde{l}_{\partial}(\ve{u}_i;\xi)\notag\\ 
&= n \ve{b}_{\partial}^{\T} (\ve{\theta}_0,\ve{\theta}_1;\xi)\Big( \ve{\hat{\mu}_{\ve{\phi}}}(\ve{U}_n) -  \ve{\mu}_{\ve{\phi}}\big( \ve{\tilde{\theta}}(\xi) \big)  \Big)
\end{align}
or
\begin{align}\label{lrr:exp:approx}
\tilde{l}(\ve{U}_n;\xi) &= n \ve{b}^{\T} (\ve{\theta}_0,\ve{\theta}_1;\xi)\Big( \ve{\hat{\mu}_{\ve{\phi}}}(\ve{U}_n) -  \ve{\mu}_{\ve{\phi}}\big( \ve{\tilde{\theta}}(\xi) \big)  \Big),
\end{align}
to the decision thresholds \eqref{sprt:thresh:low} and \eqref{sprt:thresh:up}. Note that with such an approach \eqref{LLR:hyperplane:der} or \eqref{LLR:hyperplane} are determined offline such that during run-time, the sequential test consists of updating \eqref{empirical:mean:suff:stat}, taking a vector difference, and calculating an inner product. The decision-making latency \eqref{def:asn} can be assessed using \eqref{ASN0} and \eqref{ASN1} with the mean of the ALLR
\begin{align}\label{sprt:performance:exp:approx:der}
\tilde{\mu}_{\partial,i}(\xi)&=\exdi{\ve{u};\ve{\theta}_i}{ \tilde{l}_{\partial}(\ve{u};\xi) }\notag\\
&=\ve{b}_{\partial}^{\T}(\ve{\theta}_0,\ve{\theta}_1;\xi)\Big( \ve{\mu}_{\ve{\phi}}(\ve{\theta}_i) -\ve{\mu}_{\ve{\phi}}\big( \ve{\tilde{\theta}}(\xi) \big) \Big),
\end{align}
or
\begin{align}\label{sprt:performance:exp:approx}
\tilde{\mu}_{i}(\xi)&=\ve{b}^{\T}(\ve{\theta}_0,\ve{\theta}_1;\xi)\Big( \ve{\mu}_{\ve{\phi}}(\ve{\theta}_i) -\ve{\mu}_{\ve{\phi}}\big( \ve{\tilde{\theta}}(\xi) \big) \Big).
\end{align}
Further, the variance of the ALLR is
\begin{align}
\tilde{\sigma}_{\partial,i}^2(\xi)&=\exdi{\ve{u};\ve{\theta}_i}{ \Big( \tilde{l}_{\partial}(\ve{u};\xi) - \tilde{\mu}_{\partial,i}(\xi)\Big)^2  }\notag\\
&= \ve{b}_{\partial}^{\T}(\ve{\theta}_0,\ve{\theta}_1;\xi) \ve{R}_{\ve{\phi}}(\ve{\theta}_i) \ve{b}_{\partial}(\ve{\theta}_0,\ve{\theta}_1;\xi),
\end{align}
or
\begin{align}
\tilde{\sigma}_{i}^2(\xi)=\ve{b}^{\T}(\ve{\theta}_0,\ve{\theta}_1;\xi) \ve{R}_{\ve{\phi}}(\ve{\theta}_i) \ve{b}(\ve{\theta}_0,\ve{\theta}_1;\xi).
\end{align}
\subsection{Tuning of the Probabilistic Linearization Model}
The choice of the probabilistic linearization model $p_{\ve{u}}(\ve{u};\ve{\tilde{\theta}}(\xi))$ impacts the accuracy of \eqref{lrt:approx:derivative} and \eqref{lrt:approx}. While one can use $\xi=\frac{1}{2}$, we propose a heuristic method to adapt the approximation parameter $\xi$. The idea is as follows: Instead of using the \textit{geometric} midpoint between $\ve{\theta_0}$ and $\ve{\theta}_1$, we use the \textit{statistical} midpoint. That is, we tune $\xi$ such that the approximated test statistic $\tilde{l}(\ve{u})$ admits the same properties under both hypotheses. More precisely, we consider the standardized drift
\begin{equation}\label{definition:drift}
    \tilde{d}_i(\xi) = \frac{\lvert \tilde{\mu}_i(\xi) \rvert}{\tilde{\sigma}^{\rho}_i(\xi)}, 
\end{equation}
where $\rho > 0$ can be chosen freely. Note that \eqref{definition:drift} is closely related to the error probabilities of the underlying statistical test: for $d_i = 0$, the test does not admit a drift towards any threshold such that it decides randomly; for $d_i \to \infty$, the mean of the LLR dominates the variance making the test decide correctly for $\mathcal{H}_i$ with probability one. To balance the decision-making performances under both hypotheses, it needs to hold that $\tilde{d}_0(\xi)\approx\tilde{d}_1(\xi)$. Therefore, we define the ratio
\begin{equation}
    \tilde{\nu}(\xi) = \frac{ \lvert \tilde{\mu}_1(\xi)| \tilde{\sigma}^{\rho}_0(\xi) }{\lvert \tilde{\mu}_0(\xi) \rvert \tilde{\sigma}^{\rho}_1(\xi) }
\end{equation}
and choose the linearization parameter $\xi$ such that the difference between the drifts is minimized
\begin{equation}\label{tuning:balance}
    \xi^\ast = \argmin_{\xi\in[0;1]}\, \big(\tilde{\nu}(\xi)-1\big)^2.
\end{equation}
This approach results in high-quality LLR approximations, as illustrated by examples in Sec.~\ref{sec:accuracy}.
\subsection{Approximations for the Kullback--Leibler Divergence}
Within the exponential family \eqref{def:exp:family}, the results \eqref{sprt:performance:exp:approx:der} and \eqref{sprt:performance:exp:approx} imply approximations for the Kullback--Leibler divergence
\begin{align}\label{kl:01:approx:der}
D(p_{\ve{u};\ve{\theta}_0}||p_{\ve{u};\ve{\theta}_1})&\approx -(\ve{\theta}_1-\ve{\theta}_0)^{\T} \bigg( \frac{\partial \ve{\mu}_{\ve{\phi}}\big( \ve{\tilde{\theta}}(\xi) \big) }{\partial \ve{\theta}}  \bigg)^{\T}\notag\\
&\cdot\ve{R}_{\ve{\phi}}^{-1}\big( \ve{\tilde{\theta}}(\xi) \big)\Big( \ve{\mu}_{\ve{\phi}}(\ve{\theta}_0) -\ve{\mu}_{\ve{\phi}}\big( \ve{\tilde{\theta}}(\xi) \big) \Big),\\
\label{kl:10:approx:der}
D(p_{\ve{u};\ve{\theta}_1}||p_{\ve{u};\ve{\theta}_0})&\approx (\ve{\theta}_1-\ve{\theta}_0)^{\T} \bigg( \frac{\partial \ve{\mu}_{\ve{\phi}}\big( \ve{\tilde{\theta}}(\xi) \big) }{\partial \ve{\theta}}  \bigg)^{\T}\notag\\
&\cdot\ve{R}_{\ve{\phi}}^{-1}\big( \ve{\tilde{\theta}}(\xi) \big)\Big( \ve{\mu}_{\ve{\phi}}(\ve{\theta}_1) -\ve{\mu}_{\ve{\phi}}\big( \ve{\tilde{\theta}}(\xi) \big) \Big)
\end{align}
or
\begin{align}\label{kl:01:approx}
D(p_{\ve{u};\ve{\theta}_0}||p_{\ve{u};\ve{\theta}_1})&\approx
-\big(\ve{\mu}_{\ve{\phi}}(\ve{\theta}_1)-\ve{\mu}_{\ve{\phi}}(\ve{\theta}_0)\big)^{\T} \notag\\
&\cdot\ve{R}_{\ve{\phi}}^{-1}\big( \ve{\tilde{\theta}}(\xi) \big)\Big( \ve{\mu}_{\ve{\phi}}(\ve{\theta}_0) -\ve{\mu}_{\ve{\phi}}\big( \ve{\tilde{\theta}}(\xi) \big) \Big),\\
\label{kl:10:approx}
D(p_{\ve{u};\ve{\theta}_1}||p_{\ve{u};\ve{\theta}_0})&\approx 
\big(\ve{\mu}_{\ve{\phi}}(\ve{\theta}_1)-\ve{\mu}_{\ve{\phi}}(\ve{\theta}_0)\big)^{\T}\notag\\
&\cdot\ve{R}_{\ve{\phi}}^{-1}\big( \ve{\tilde{\theta}}(\xi) \big)\Big( \ve{\mu}_{\ve{\phi}}(\ve{\theta}_1) -\ve{\mu}_{\ve{\phi}}\big( \ve{\tilde{\theta}}(\xi) \big) \Big).
\end{align}
With the Fisher information matrix for exponential family distributions \eqref{def:exp:family} being characterized by \cite{Stein_FisherBound}
\begin{align}\label{def:fisher:matrix:exp:family}
\ve{F}(\ve{\theta}) &= \exdi{\ve{u};\ve{\theta}}{ \bigg( \frac{\partial \ln p_{\ve{u}}(\ve{u};\ve{\theta})}{\partial \ve{\theta}} \bigg)^{\T} \frac{\partial \ln p_{\ve{u}}(\ve{u};\ve{\theta})}{\partial \ve{\theta}} }\notag\\
&=\bigg( \frac{\partial \ve{\mu}_{\ve{\phi}}\big( \ve{\theta} \big) }{\partial \ve{\theta}}  \bigg)^{\T} \ve{R}_{\ve{\phi}}^{-1}\big( \ve{\theta} \big)\frac{\partial \ve{\mu}_{\ve{\phi}}\big( \ve{\theta} \big) }{\partial \ve{\theta}},
\end{align}
applying forward \eqref{dif:approx:forward:vec} and backward \eqref{dif:approx:backward:vec} approximations to \eqref{kl:01:approx:der} and \eqref{kl:10:approx:der}, one also obtains
\begin{align}\label{kl:01:approx:der:fish}
D(p_{\ve{u};\ve{\theta}_0}||p_{\ve{u};\ve{\theta}_1})&\approx (\ve{\theta}_1-\ve{\theta}_0)^{\T} \ve{F}\big( \ve{\tilde{\theta}}(\xi) \big)\big(\ve{\tilde{\theta}}(\xi)- \ve{\theta}_0\big),\\
\label{kl:10:approx:der:fish}
D(p_{\ve{u};\ve{\theta}_1}||p_{\ve{u};\ve{\theta}_0})&\approx (\ve{\theta}_1-\ve{\theta}_0)^{\T} \ve{F}\big( \ve{\tilde{\theta}}(\xi) \big)\big( \ve{\theta}_1 -\ve{\tilde{\theta}}(\xi) \big).
\end{align}
These Kullback--Leibler divergence approximations are reminiscent of expressions found in the literature \cite[pp. 85-86]{TartakovskyBook}
\begin{align}\label{kl:01:approx:lit}
D(p_{\ve{u};\ve{\theta}_0}||p_{\ve{u};\ve{\theta}_1})&\approx - \frac{1}{2} (\ve{\theta}_1-\ve{\theta}_0)^{\T} \ve{F}\big( \ve{\theta}_0 \big)\big( \ve{\theta}_1-\ve{\theta}_0 \big)\notag\\
&=\hat{D}(p_{\ve{u};\ve{\theta}_0}||p_{\ve{u};\ve{\theta}_1}),\\
\label{kl:10:approx:lit}
D(p_{\ve{u};\ve{\theta}_1}||p_{\ve{u};\ve{\theta}_0})&\approx \frac{1}{2} (\ve{\theta}_1-\ve{\theta}_0)^{\T} \ve{F}\big( \ve{\theta}_1 \big)\big( \ve{\theta}_1-\ve{\theta}_0 \big)\notag\\
&=\hat{D}(p_{\ve{u};\ve{\theta}_1}||p_{\ve{u};\ve{\theta}_0}),
\end{align}
while having a flexible probabilistic linearization point $\ve{\tilde{\theta}}(\xi)$. Note that the approximations \eqref{kl:01:approx:lit} and \eqref{kl:10:approx:lit} are derived in \cite{TartakovskyBook} for the special case of a natural exponential family, i.e., under the restriction $\ve{\beta}(\ve{\theta})=\ve{\theta}$. A potential disadvantage of \eqref{kl:01:approx:der:fish} and \eqref{kl:10:approx:der:fish} is that access to the Fisher information matrix \eqref{def:fisher:matrix:exp:family} of the exponential family model $p_{\ve{u}}(\ve{u};\ve{\tilde{\theta}}(\xi))$ is required.
\subsection{Controlling the Statistical Complexity of the ALLR}
While \eqref{lrt:approx:derivative} and \eqref{lrt:approx} enable running and analyzing a likelihood-oriented test via \eqref{exp:family:mean} and \eqref{exp:family:covariance} without direct access to the likelihood ratio \eqref{def:llr:exp:fam}, they do not per se solve its intractability with a large number of sufficient statistics $C$ (as with multivariate binary distributions). Under such circumstances, replacing the data model by an auxiliary version \cite{Stein_FisherBound}
\begin{align}\label{def:exp:rep}
\tilde{p}_{\ve{u}}(\ve{u};\ve{\theta})=\exp{\ve{\tilde{\beta}}^{\T}(\ve{\theta}) \ve{\tilde{\phi}}(\ve{u}) - \tilde{\lambda}(\ve{\theta})+\tilde{\nu}(\ve{u})},
\end{align}
allows to control the ALLR complexity. In \eqref{def:exp:rep}, $\ve{\tilde{\phi}}(\ve{u}) \colon \ve{\mathcal{U}} \to\fieldR^{\tilde{C}}$ is a subset ($\tilde{C}<C$) of the sufficient statistics in the model \eqref{def:exp:family}, i.e., $\ve{\phi}(\ve{u})=\begin{bmatrix} \ve{\tilde{\phi}}^{\T}(\ve{u}) &\ve{\phi}'^{\T}(\ve{u}) \end{bmatrix}^{\T}$ while mean 
\begin{align}\label{auxiliary:mean}
\ve{\mu}_{\ve{\tilde{\phi}}}(\ve{\theta}) = \exdi{\ve{u};\ve{\theta}}{ \ve{\tilde{\phi}}(\ve{u}) }
\end{align}
and covariance
\begin{align}\label{auxiliary:covariance}
\ve{R}_{\ve{\tilde{\phi}}}(\ve{\theta})&=\exdi{\ve{u};\ve{\theta}}{ \big( \ve{\tilde{\phi}}(\ve{u}) - \ve{\mu}_{\ve{\tilde{\phi}}}(\ve{\theta}) \big) \big(\ve{\tilde{\phi}}(\ve{u})-\ve{\mu}_{\ve{\tilde{\phi}}}(\ve{\theta}) \big)^{\T}}
\end{align}
are equivalent when formed with respect to \eqref{def:exp:rep} or \eqref{def:exp:family}. It can then be shown that \cite{Stein_FisherBound} 
\begin{align}\label{pessimistic:fisher}
\ve{F}(\ve{\theta}) \succeq \ve{\tilde{F}}(\ve{\theta}),
\end{align}
where $\ve{\tilde{F}}(\ve{\theta})$ is the Fisher information matrix \eqref{def:fisher:matrix:exp:family} of the auxiliary data model \eqref{def:exp:rep}. Due to the matrix inequality \eqref{pessimistic:fisher} and the quadratic form of \eqref{kl:01:approx:der:fish} and \eqref{kl:10:approx:der:fish}, \eqref{def:exp:rep} constitutes a conservative modification of the ALLRs \eqref{lrt:approx:derivative} and \eqref{lrt:approx}. 

When evaluating \eqref{auxiliary:mean} and \eqref{auxiliary:covariance} for the hard-limited version of \eqref{digital:model:multivariate:gauss}, we use an auxiliary model \eqref{def:exp:rep} with statistics \cite{Stein_FisherBound}
\begin{align}\label{aux:statistics:quantizer}
\ve{\tilde{\phi}}(\ve{z})&=\ve{\Phi}\vec{\ve{z} \ve{z}^{\T}},
\end{align}
where $\ve{\Phi}\in [0; 1]^{\tilde{C} \times (MK)^2}$ is an elimination matrix canceling the duplicate and constant statistics on $\ve{z} \ve{z}^{\T}$. This reduces the binary data model to a quadratic exponential binary distribution \cite{Cox94} where the statistical complexity is $\tilde{C}=\frac{MK}{2}(MK - 1)$. For the calculation of the mean \eqref{auxiliary:mean} and the covariance \eqref{auxiliary:covariance} with \eqref{aux:statistics:quantizer}, we proceed as described in \cite{Stein16_WSA} and use the classic arcsine law \cite[p. 284]{ThomasBook} together with recent results for the quadrivariate orthant probabilities \cite{Sinn11}. Note that, through the concept of a reduced distribution \eqref{def:exp:rep}, with \eqref{exp:family:mean} and \eqref{exp:family:covariance} for a subset of sufficient statistics at hand, the presented framework is also applicable to other sensor data models like receiver systems with higher A/D resolution.
\section{Accuracy of the LLR Approximation}\label{sec:accuracy}
\subsection{Digital Data Model and Error Measures}
To analyze the accuracy of \eqref{lrt:approx:derivative} and \eqref{lrt:approx}, we consider a $K$-variate Gaussian model with scalar parameter $\theta\in\fieldR$
\begin{align}\label{approx:accuracy:Gauss:example:dist}
p_{\ve{y}}(\ve{y};\theta)=\frac{\exp{-\frac{1}{2} \ve{y}^{\T} \ve{R}_{\ve{y}}^{-1}(\theta) \ve{y}}}{ \sqrt{(2\pi)^{K} \det{(\ve{R}_{\ve{y}}(\theta))}} }.
\end{align}
Note, that by
\begin{align}
\ve{y}^{\T} \ve{R}_{\ve{y}}^{-1}(\theta) \ve{y}&=\tr{\ve{R}_{\ve{y}}^{-1}(\theta) \ve{y} \ve{y}^{\T}}\notag\\
&=\vecT{\ve{R}_{\ve{y}}^{-1}(\theta)}\vec{ \ve{y} \ve{y}^{\T} },
\end{align}
\eqref{approx:accuracy:Gauss:example:dist} can be factorized along \eqref{def:exp:family} by using $\nu(\ve{y}) = 0$ and
\begin{align}
\ve{\beta}(\theta)&=-\frac{1}{2}\vec{\ve{R}_{\ve{y}}^{-1}(\theta)},\label{natural:paramters:Gauss}\\
\ve{\phi}(\ve{y})&=\vec{\ve{y}\ve{y}^{\T}},\label{sufficient:statistics:Gauss}\\
\lambda(\theta)&=\frac{1}{2}\ln \det{(\ve{R}_{\ve{y}}(\theta))} + \frac{K}{2}\ln 2\pi.
\end{align}
Additionally, with \eqref{sufficient:statistics:Gauss}, one obtains
\begin{align}
\ve{\mu}_{\ve{\phi}}(\theta) &=\vec{\ve{R}_{\ve{y}}(\theta)},\\
\ve{R}_{\ve{\phi}}(\theta) &= 2 \big( \ve{R}_{\ve{y}}(\theta) \otimes \ve{R}_{\ve{y}}(\theta) \big).
\end{align}
The exact expected LLRs between two distributions \eqref{approx:accuracy:Gauss:example:dist} are
\begin{align}\label{kl:divergence:gaussian:h0}
{\mu}_0&=\frac{1}{2}\ln  \frac{ \det{ \big( \ve{R}_{\ve{y}}({\theta}_0) \big)}  }{ \det{ \big( \ve{R}_{\ve{y}}({\theta}_1)}  \big) } +\frac{K}{2}-\frac{1}{2} \tr{ \ve{R}^{-1}_{\ve{y}}({\theta}_1) \ve{R}_{\ve{y}}({\theta}_0) },\\
\label{kl:divergence:gaussian:h1}
{\mu}_1&=\frac{1}{2}\ln  \frac{ \det{ \big( \ve{R}_{\ve{y}}({\theta}_0) \big)}  }{ \det{ \big( \ve{R}_{\ve{y}}({\theta}_1)}  \big) } -\frac{K}{2}+\frac{1}{2} \tr{ \ve{R}^{-1}_{\ve{y}}({\theta}_0) \ve{R}_{\ve{y}}({\theta}_1) }.
\end{align}
Therefore, under the Gaussian distribution \eqref{approx:accuracy:Gauss:example:dist}, we can evaluate the relative approximation errors
\begin{align}\label{def:rel:approx:error}
\tilde{\epsilon}_i (\xi)&= \frac{  \big| \tilde{\mu}_i(\xi) \big| - \big| {\mu}_i \big|  }{ \big| {\mu}_i \big| },
\end{align}
and compare them to \eqref{kl:01:approx:lit} and \eqref{kl:10:approx:lit} through the expression
\begin{align}\label{def:rel:old:approx:error}
\hat{\epsilon}_i &= \frac{  \hat{D}(p_{\ve{u};\ve{\theta}_i}||p_{\ve{u};\ve{\theta}_j}) - D(p_{\ve{u};\ve{\theta}_i}||p_{\ve{u};\ve{\theta}_j})  }{ D(p_{\ve{u};\ve{\theta}_i}||p_{\ve{u};\ve{\theta}_j}) }, \quad i \neq j.
\end{align}
\subsection{Application - Sampling Random Gaussian Processes}
To connect the multivariate Gaussian model \eqref{approx:accuracy:Gauss:example:dist} to a practical sensing application, we assume that it models the digital data stream obtained by sampling \eqref{receive:signal:analog} with $M=1$ via an ideal A/D converter featuring $\infty$-bit amplitude resolution. The continuous-time Gaussian random process $y(t)\in\fieldR$ is assumed to be the superposition of a band-limited source signal $\breve{s}(t)\in\fieldR$ and white noise $\breve{\eta}(t)$ after preprocessing with an ideal low-pass filter $h(t;B)$ of bandwidth $B=B_y$, i.e.,
\begin{align}
y(t)=h(t;B_y)\ast \big(\breve{s}(t) + \breve{\eta}(t)\big).
\end{align}
The band-limited source $\breve{s}(t)$ features power spectral density $\Psi_s(\omega)=\Psi_s, \omega\in [ -\Omega_s;  \Omega_s]$ with $\Omega_s \leq \Omega_y$, and $\Psi_s(\omega)=0$ elsewhere. Consequently, the source passes unaffected
\begin{align}
s(t)=h(t;B_y)\ast \breve{s}(t)=\breve{s}(t).
\end{align}
The noise $\breve{\eta}(t)$ has constant power spectral density $\Psi_\eta(\omega)=\Psi_\eta$ on $\omega\in [ -\Omega_\eta;  \Omega_\eta]$ with $\Omega_\eta \gg \Omega_y$. Consequently, 
\begin{align}
\eta(t)=h(t;B_y)\ast \breve{\eta}(t)\neq\breve{\eta}(t).
\end{align}
Note that, by the Wiener--Khintchine theorem \cite{Khintchine34}, the auto-correlation function of a band-limited random process $u(t)$ (bandwidth $B_u$ or $\Omega_u$) with power spectral density $\Psi_u$ is
\begin{align}
r_u(t)&=\exdi{u}{u(\tau) u(\tau-t)}%\notag\\
%&=\frac{1}{2\pi}\int_{-\infty}^{\infty} \Psi_u(\omega) \e^{{\rm j} \omega t} {\rm d}\omega\notag\\
= \frac{1}{2\pi} \int_{-\Omega_u}^{\Omega_u} \Psi_u(\omega) \e^{{\rm j} \omega t} {\rm d}\omega\notag\\
&=\frac{\Psi_u}{\pi} \frac{\sin{(2 \pi B_u t)}}{t}%\notag\\
=2 B_u \Psi_u \sinc{2 B_u t}.
\end{align}
With the low-pass filter $h(t;B_y)$ including a gain-control factor $\tfrac{1}{\sqrt{2B_y\Psi_\eta}}$ and sampling $K$ times at a rate $f_{\text{T}}=\tfrac{1}{T}=2B_y$
\begin{align}\label{gaussian:sampling:covariance}
\ve{R}_{\ve{y}}(\theta)&= \ve{R}_{\ve{s}}(\theta)+ \ve{R}_{\ve{\eta}}%\notag\\
=\frac{\theta}{\kappa} \ve{\Sigma}(\kappa)+\ve{I},
\end{align}
where the signal-to-noise ratio (SNR) is
\begin{align}\label{def:snr:freq}
\theta=\text{SNR}=\frac{\Psi_s}{\Psi_\eta}.
\end{align}
With a source oversampling factor $\kappa\in\fieldR,\kappa=\tfrac{B_y}{B_s} \geq 1$, the source covariance matrix has entries
\begin{align}\label{source:covariance:entries}
[ \ve{\Sigma}(\kappa)]_{ij}=\sinc{\frac{|i-j|}{\kappa}}, \quad i,j=1,\ldots,K.
\end{align}
Note that, in contrast to the source covariance $\ve{R}_{\ve{s}}(\theta)$, in \eqref{gaussian:sampling:covariance} the noise covariance $\ve{R}_{\ve{\eta}}$ is the identity matrix (temporally white noise) irrespective of $\kappa$ as $f_{\text{T}}=2B_y=\kappa 2B_s$. 
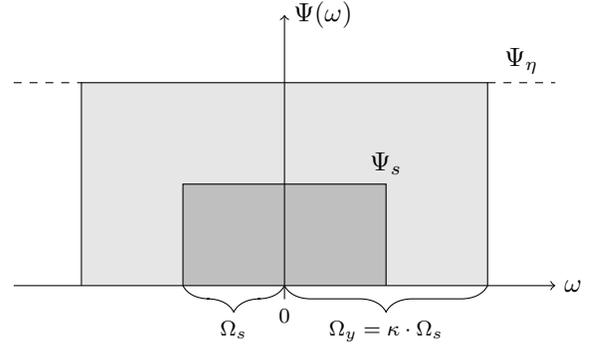
\begin{figure}[!ht]
    \centering
      \begin{tikzpicture}[scale=0.90]

  	% Sensor noise
  	\draw[fill=gray!20] (-3,0) --(-3,3) -- (3,3) -- (3,0);
  	\draw [decorate,decoration={brace,amplitude=10pt,mirror}] (0,0) -- (3,0) node [black,midway,below,yshift=-10pt] {\footnotesize $\Omega_y=\kappa \cdot \Omega_s$};
  	\draw (3.5,3) node[above] {$\Psi_\eta$};
  
  	\draw[style=dashed] (-4,3) --(-3,3);
  	\draw[style=dashed] (3,3) --(4,3);
  
  	% Source 1
  	\draw[fill=gray!50] (-1.5,0) -- (-1.5,1.5) -- (1.5,1.5) -- (1.5,0);
  	\draw [decorate,decoration={brace,amplitude=10pt,mirror}] (-1.5,0) -- (0,0) node [black,midway,below,yshift=-10pt] {\footnotesize $\Omega_s$};
  	\draw (1.5,1.5) node[above] {$\Psi_s$};
  
  	\draw[->] (-4,0) -- (4,0) node[right] {$\omega$};
  	\draw[->] (0,-0.2) -- (0,4) node[right] {$\Psi(\omega)$};
	
	\draw (0,-0.2) node[below] {\footnotesize $0$};
  
\end{tikzpicture}
    \caption{Power Spectral Densities and Signal Bandwidths ($\kappa=2$, $\text{SNR}=\SI{-3.0}{\decibel}$)}
    \label{example:frequency}
\end{figure}

For illustration, Fig.~\ref{example:frequency} visualizes an exemplary sensing situation with oversampling $\kappa=2$ and $\text{SNR}=\SI{-3.0}{\decibel}$ in the frequency domain. Note, that we define \eqref{def:snr:freq} independently of $h(t;B_y)$ and $\kappa$ to ensure that the design of the analog preprocessing does not affect the physical SNR.
\subsection{Results - Approximation Accuracy}
\begin{figure}[!ht]
    \centering
    \begin{subfigure}[t]{0.5\textwidth}
        \centering
        \begin{tikzpicture}[scale=0.95]

  	\begin{axis}[ylabel=$\text{Approximation Error $\epsilon_i$}$,
  			xlabel=$\text{Signal-to-Noise Ratio $\theta_1$ [dB]}$ ,
			grid,
			ymin=-0.4,
			ymax=0.4,
			xmin=-20,
			xmax=0,
			legend columns=3,
			legend pos=south west,
			height = 190pt,
			width = 240pt]
				
			\addplot[blue, style=dotted, line width=0.75pt,smooth, every mark/.append style={solid}, mark=square*, mark phase = 0, mark repeat=4] table[x index=0, y index=5]{Data/ALLR_Accuracy_GaussianSampling_kappa_2_N_10_SNR0_-20.txt};
			\addlegendentry{$\hat{\epsilon}_0$}
					
			\addplot[blue, style=dashed, line width=0.75pt,smooth, every mark/.append style={solid}, mark=square*, mark phase = 0, mark repeat=4] table[x index=0, y index=1]{Data/ALLR_Accuracy_GaussianSampling_kappa_2_N_10_SNR0_-20.txt};
			\addlegendentry{$\tilde{\epsilon}_0(\frac{1}{2})$}
			
			\addplot[blue, style=solid, line width=0.75pt,smooth, every mark/.append style={solid}, mark=square*, mark phase = 0, mark repeat=4] table[x index=0, y index=3]{Data/ALLR_Accuracy_GaussianSampling_kappa_2_N_10_SNR0_-20.txt};
			\addlegendentry{$\tilde{\epsilon}_0(\xi^\ast)$}
			
			\addplot[red, style=dotted, line width=0.75pt,smooth, every mark/.append style={solid}, mark=*, mark phase = 3, mark repeat=4] table[x index=0, y index=6]{Data/ALLR_Accuracy_GaussianSampling_kappa_2_N_10_SNR0_-20.txt};
			\addlegendentry{$\hat{\epsilon}_1$}
			
		 	\addplot[red, style=dashed, line width=0.75pt,smooth, every mark/.append style={solid}, mark=*, mark phase = 3, mark repeat=4] table[x index=0, y index=2]{Data/ALLR_Accuracy_GaussianSampling_kappa_2_N_10_SNR0_-20.txt};
			\addlegendentry{$\tilde{\epsilon}_1(\frac{1}{2})$}

		 	\addplot[red, style=solid, line width=0.75pt,smooth, every mark/.append style={solid}, mark=*, mark phase = 3, mark repeat=4] table[x index=0, y index=4]{Data/ALLR_Accuracy_GaussianSampling_kappa_2_N_10_SNR0_-20.txt};
			\addlegendentry{$\tilde{\epsilon}_1(\xi^\ast)$}
								
	\end{axis}
	
  \end{tikzpicture}
        \caption{Approximation Quality}
        \label{fig:approx:quality}
    \end{subfigure}
    \begin{subfigure}[t]{0.5\textwidth}
        \centering
          \begin{tikzpicture}[scale=0.95]

  	\begin{axis}[ylabel=$\text{Approximation Parameter $\xi$}$,
  			xlabel=$\text{Signal-to-Noise Ratio $\theta_1$ [dB]}$ ,
			grid,
			ymin=0.48,
			ymax=0.60,
			xmin=-20,
			xmax=0,
			legend pos=south east,
			height = 150pt,
			width = 240pt]
						
			\addplot[black, line width=0.75pt, smooth] table[x index=0, y index=7]{Data/ALLR_Accuracy_GaussianSampling_kappa_2_N_10_SNR0_-20.txt};
			\addlegendentry{$\xi^\ast$}
															
	\end{axis}
	
\end{tikzpicture}
        \caption{Probabilistic Linearization Model}
        \label{fig:approx:quality:tuning}
    \end{subfigure}%
    \caption{ALLR - Approximation Quality and Tuning ($K=10, \kappa=2, \theta_0=\SI{-20}{\decibel}$)}
\end{figure}
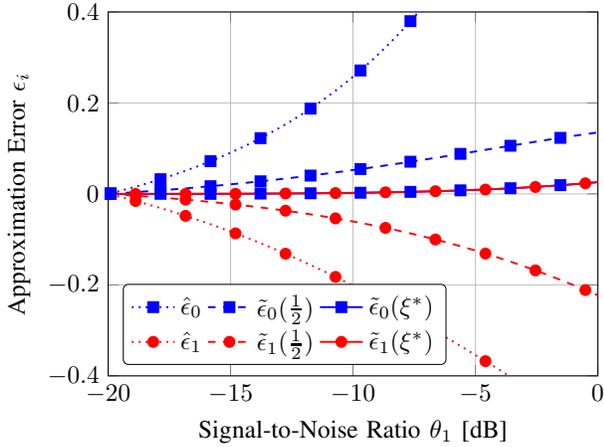
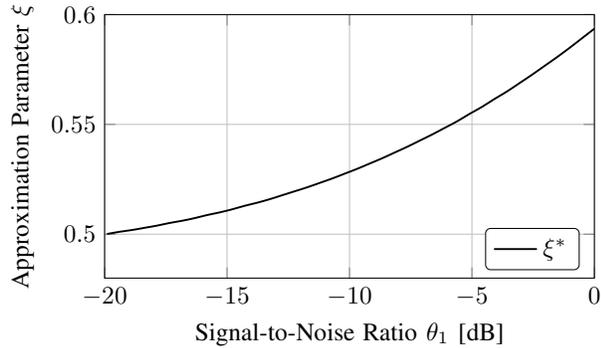

Fig.~\ref{fig:approx:quality} shows the relative errors \eqref{def:rel:approx:error} and \eqref{def:rel:old:approx:error} for a setting with $K=10, \kappa=2,  \theta_0=\SI{-20}{\decibel}$ as a function of $\theta_1$. It can be observed that the error $|\hat{\epsilon}_i|$ increases quickly with the distance $\theta_1-\theta_0$ and beyond $\theta_1=\SI{-10}{\decibel}$ exceeds $\SI{27.1}{\percent}$. In contrast, for all considered SNR values, the error $|\epsilon_i(\frac{1}{2})|$ is not larger than $\SI{22.3}{\percent}$. Using \eqref{tuning:balance} with $\rho=\frac{2}{3}$, the error $|\epsilon_i(\xi^\ast)|$ is below $\SI{2.6}{\percent}$ over the entire depicted SNR range. Fig. \ref{fig:approx:quality:tuning} shows $\xi^\ast$ resulting from \eqref{tuning:balance} with $\rho=\frac{2}{3}$.

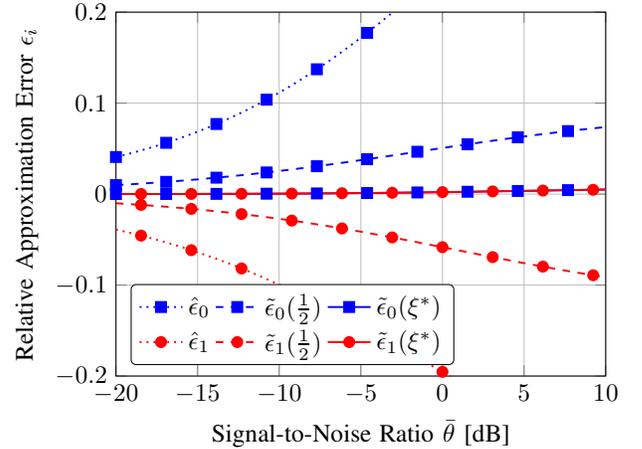
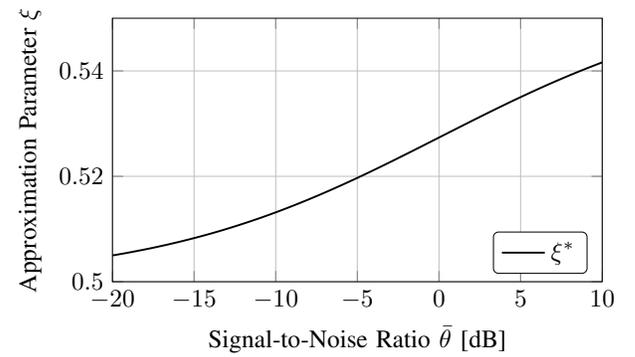
\begin{figure}[!ht]
    \centering
    \begin{subfigure}[t]{0.5\textwidth}
        \centering
          \begin{tikzpicture}[scale=0.95]

  	\begin{axis}[ylabel=$\text{Relative Approximation Error $\epsilon_i$}$,
  			xlabel=$\text{Signal-to-Noise Ratio $\bar{\theta}$ [dB]}$ ,
			grid,
			ymin=-0.2,
			ymax=0.2,
			xmin=-20,
			xmax=10,
			legend columns=3,
			legend pos=south west,
			height = 190pt,
			width = 240pt]
						
			\addplot[blue, style=dotted, line width=0.75pt,smooth, every mark/.append style={solid}, mark=square*, mark phase = 0, mark repeat=4] table[x index=0, y index=7]{Data/ALLR_Accuracy_GaussianSampling_sweep_kappa_2_N_10_SNRdelta_1.5.txt};
			\addlegendentry{$\hat{\epsilon}_0$}
					
			\addplot[blue, style=dashed, line width=0.75pt,smooth, every mark/.append style={solid}, mark=square*, mark phase = 0, mark repeat=4] table[x index=0, y index=1]{Data/ALLR_Accuracy_GaussianSampling_sweep_kappa_2_N_10_SNRdelta_1.5.txt};
			\addlegendentry{$\tilde{\epsilon}_0(\frac{1}{2})$}
			
			\addplot[blue, style=solid, line width=0.75pt,smooth, every mark/.append style={solid}, mark=square*, mark phase = 0, mark repeat=4] table[x index=0, y index=3]{Data/ALLR_Accuracy_GaussianSampling_sweep_kappa_2_N_10_SNRdelta_1.5.txt};
			\addlegendentry{$\tilde{\epsilon}_0(\xi^\ast)$}
			
			\addplot[red, style=dotted, line width=0.75pt,smooth, every mark/.append style={solid}, mark=*, mark phase = 3, mark repeat=4] table[x index=0, y index=8]{Data/ALLR_Accuracy_GaussianSampling_sweep_kappa_2_N_10_SNRdelta_1.5.txt};
			\addlegendentry{$\hat{\epsilon}_1$}
			
		 	\addplot[red, style=dashed, line width=0.75pt,smooth, every mark/.append style={solid}, mark=*, mark phase = 3, mark repeat=4] table[x index=0, y index=2]{Data/ALLR_Accuracy_GaussianSampling_sweep_kappa_2_N_10_SNRdelta_1.5.txt};
			\addlegendentry{$\tilde{\epsilon}_1(\frac{1}{2})$}

		 	\addplot[red, style=solid, line width=0.75pt,smooth, every mark/.append style={solid}, mark=*, mark phase = 3, mark repeat=4] table[x index=0, y index=4]{Data/ALLR_Accuracy_GaussianSampling_sweep_kappa_2_N_10_SNRdelta_1.5.txt};
			\addlegendentry{$\tilde{\epsilon}_1(\xi^\ast)$}
												
	\end{axis}
	
\end{tikzpicture}
        \caption{Approximation Quality}
        \label{fig:approx:quality:small}
    \end{subfigure}
    \begin{subfigure}[t]{0.5\textwidth}
        \centering
        \begin{tikzpicture}[scale=0.95]

  	\begin{axis}[ylabel=$\text{Approximation Parameter $\xi$}$,
  			xlabel=$\text{Signal-to-Noise Ratio $\bar{\theta}$ [dB]}$ ,
			grid,
			ymin=0.5,
			ymax=0.55,
			xmin=-20,
			xmax=10,
			legend pos=south east,
			height = 150pt,
			width = 240pt]
						
			\addplot[black, line width=0.75pt, smooth] table[x index=0, y index=9]{Data/ALLR_Accuracy_GaussianSampling_sweep_kappa_2_N_10_SNRdelta_1.5.txt};
			\addlegendentry{$\xi^\ast$}
															
	\end{axis}
	
\end{tikzpicture}
        \caption{Probabilistic Linearization Model}
        \label{fig:approx:quality:small:tuning}
    \end{subfigure}%
\caption{ALLR - Approximation Quality and Tuning ($K=10, \kappa=2, \Delta_\theta=\SI{1.5}{\decibel}$)}
\end{figure}

Note that a typical sequential task is to detect small SNR changes from the digital measurements. To asses the quality of \eqref{lrt:approx} in such a context, using the definition $\theta_0=\bar{\theta}-\Delta_\theta$ and $\theta_1=\bar{\theta}+\Delta_\theta$, in Fig.~\ref{fig:approx:quality:small}, we visualize the relative errors \eqref{def:rel:approx:error} and \eqref{def:rel:old:approx:error} for $\Delta_\theta=\SI{1.5}{\decibel}, K=10, \kappa=2$, as a function of $\bar{\theta}$. The error $|\hat{\epsilon}_i|$ exceeds $\SI{16.6}{\percent}$ for $\bar{\theta}\geq-\SI{5}{\decibel}$. The relative error $|\epsilon_i(\frac{1}{2})|$ stays below $\SI{9.2}{\percent}$ for all SNR values, while employing \eqref{tuning:balance} with $\rho=\frac{2}{3}$ results in $\xi^\ast$ shown in Fig.~\ref{fig:approx:quality:small:tuning} and an error $|\epsilon_i(\xi^\ast)|$ smaller than $\SI{0.52}{\percent}$. The results for the exemplary model \eqref{approx:accuracy:Gauss:example:dist} show that \eqref{lrt:approx} is of high quality, in particular if the probabilistic linearization $\tilde{\theta}(\xi)$ is optimized.

\section{Application - Binary Radio System Design}\label{sec:application}
Next, the impact of theoretical results like \eqref{lrt:approx:derivative} or \eqref{lrt:approx} on technical systems is illustrated by exemplary applications in radio system engineering. To emphasize the significance of binary sensing and data processing for future wireless systems at large, we use a consumer-oriented cognitive radio application with low complexity as the main focus and a safety-critical spectrum monitoring application with a distinct emphasis on high performance. For each application, a particular analog front-end is used when modeling the sensor signals \eqref{receive:signal:analog}.
\subsection{Low-Cost Cognitive Radio Communication Systems}\label{sec:cog:radio}
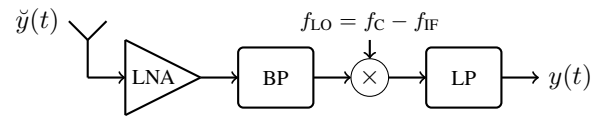
\begin{figure}[!ht]
    \centering
    \begin{tikzpicture}[scale=1]
	
		% antenna
		\draw (-1.25,0.75) node[left] {$\breve{y}(t)$};
		\draw [thick,rounded corners=2pt] (-1.0,0.5) -- (-1.0,0);
		\draw [thick,rounded corners=2pt] (-1.0,0.5) -- (-1.25,0.75);
		\draw [thick,rounded corners=2pt] (-1.0,0.5) -- (-0.75,0.75);

		% connection: antenna - LNA
		\draw [thick,rounded corners=2pt,->] (-1.0,0) -- (-0.5,0);
		
		% LNA
		\draw (-0.125,0) node {\footnotesize LNA};
		\draw [thick,rounded corners=2pt] (-0.5,-0.5) -- (-0.5,0.5);
		\draw [thick,rounded corners=2pt] (-0.5,-0.5) -- (0.5,0);
		\draw [thick,rounded corners=2pt] (-0.5,0.5) -- (0.5,0);
		
		% connection: LNA - BP
		\draw [thick,rounded corners=2pt,->] (0.5,0) -- (1.0,0);
		
		% BP
		\draw (1.5,0) node {\footnotesize BP};
		\draw [thick,rounded corners=2pt] (1.0,-0.4) rectangle (2.0,0.4);
		
		% connection: BP - LO
		\draw [thick,rounded corners=2pt,->] (2.0,0) -- (2.5,0);
		
		% connection: LO
		\draw (2.75,0) node {$\times$};
		\draw (2.75,0) circle (0.25cm);
		
		% connection: antenna - LNA
		\draw (2.75,0.5) node[above] {\footnotesize $f_{\text{LO}}=f_{\text{C}}-f_{\text{IF}}$};
		\draw [thick,rounded corners=2pt,->] (2.75,0.5) -- (2.75,0.25);
		
		% connection:  LO - LP
		\draw [thick,rounded corners=2pt,->] (3.0,0) -- (3.5,0);
		
		% LP
		\draw (4.0,0) node {\footnotesize LP};
		\draw [thick,rounded corners=2pt] (3.5,-0.4) rectangle (4.5,0.4);
		
		% connection: LP - ADC
		\draw (5.0,0) node[right] {${y}(t)$};
		\draw [thick,rounded corners=2pt,->] (4.5,0) -- (5.0,0);
		
	\end{tikzpicture}
    \caption{Superheterodyne Analog Sensor Front-End}
    \label{superhet:receiver}
\end{figure}

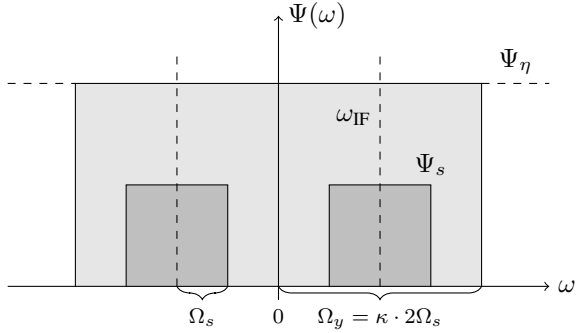
\begin{figure}[!ht]
    \centering
    \begin{tikzpicture}[scale=0.90]

  	% Sensor noise
  	\draw[fill=gray!20] (-3,0) --(-3,3) -- (3,3) -- (3,0);
	\draw (3.5,3) node[above] {$\Psi_\eta$};
	
  	\draw [decorate,decoration={brace,amplitude=5pt,mirror}] (0,0) -- (3,0) node [black,midway,below,yshift=-5pt] {\footnotesize $\Omega_y=\kappa \cdot 2\Omega_s$};
  	
  	\draw[style=dashed] (-4,3) --(-3,3);
  	\draw[style=dashed] (3,3) --(4,3);
  
  	% Source 1
	\draw[fill=gray!50] (0.75,0) -- (0.75,1.5) -- (2.25,1.5) -- (2.25,0);
  	\draw[fill=gray!50] (-0.75,0) -- (-0.75,1.5) -- (-2.25,1.5) -- (-2.25,0);
  	
	\draw [decorate,decoration={brace,amplitude=5pt,mirror}] (-1.5,0) -- (-0.75,0) node [black,midway,below,yshift=-5pt] {\footnotesize $\Omega_s$};
  	\draw (2.25,1.5) node[above] {$\Psi_s$};
	
	\draw[style=dashed] (1.5,0) --(1.5,3.5);
	\draw[style=dashed] (-1.5,0) --(-1.5,3.5);
	\draw (1.5,2.5) node[left] {$\omega_{\text{IF}}$};
  
  	\draw[->] (-4,0) -- (4,0) node[right] {$\omega$};
  	\draw[->] (0,-0.2) -- (0,4) node[right] {$\Psi(\omega)$};
	
	\draw (0,-0.2) node[below] {\footnotesize $0$};
  
\end{tikzpicture}
    \caption{Power Spectral Densities and Signal Bandwidths (Superheterodyne Receiver, $\kappa=4$, $\text{SNR}=\SI{-3.0}{\decibel}$)}
    \label{superhet:frequency}
\end{figure}

First, we consider a cognitive system for mobile communication. The task of the receiver is to monitor a certain part of the spectrum and to determine whether a primary transmitter is currently utilizing the radio channel. The scenario is, therefore, similar to the example considered in Sec. \ref{sec:accuracy}. However, when modeling the analog outputs \eqref{receive:signal:analog}, it is essential to take into account the front-end architecture. For cognitive radio, we assume a single superheterodyne receiver ($M=1$), as depicted in Fig.~\ref{superhet:receiver}.
The difficulty in modeling its analog output lies in the fact that the receiver does not demodulate with the carrier frequency $f_{\text{C}}$ of the transmitter. A local oscillator demodulates the received signal with $f_{\text{LO}}=f_{\text{C}}-f_{\text{IF}}$, where $f_{\text{IF}}< f_{\text{C}}$ is an intermediate frequency (IF). As a consequence, sampling is performed on one real-valued analog output
\begin{align}\label{analog:receive:signal:superheterodyne}
y(t)=x_{\text{I}}(t)\sqrt{2}\cos{(\omega_{\text{IF}}t)} - x_{\text{Q}}(t)\sqrt{2}\sin{(\omega_{\text{IF}}t)} + \eta(t),
\end{align}
where $x_{\text{I}}(t)$ and $x_{\text{Q}}(t)$ are assumed to be two jointly independent zero-mean Gaussian processes of bandwidth $B_s$ modeling the transmitter. The independent zero-mean Gaussian process $\eta(t)$ models noise with bandwidth $B_y$. Sampling $K$ times at a rate of $\frac{1}{T}=2B_y$ and with an amplitude resolution of $b=\infty$ bits, one obtains zero-mean multivariate Gaussian data with
\begin{align}\label{covariance:superheterodyne}
\ve{R}_{\ve{y}}(\theta)= \frac{\theta}{\kappa} \ve{\Sigma}(\kappa) \odot 2 \ve{W}+\ve{I},\quad \ve{R}_{\ve{y}}(\theta)\in\fieldR^{K \times K},
\end{align}
where $\theta$ is the physically received SNR as defined in \eqref{def:snr:freq}, $\ve{\Sigma}(\kappa)$ the source covariance as defined in \eqref{source:covariance:entries}, and the matrix $\ve{W}\in\fieldR^{K \times K}$ models the mixing effects at the intermediate frequency $f_{\text{IF}}$. For simplicity, we assume $f_{\text{IF}}$ to be symmetric with respect to the bandwidth $B_y$ of the low-pass (LP) filter, i.e., $f_{\text{IF}}=\frac{B_y}{2}$. Thus the entries of the mixing matrix are
\begin{align}
[ \ve{W} ]_{ij}&=\cos{(\omega_{\text{IF}}T(i-1))}\cos{(\omega_{\text{IF}}T(j-1))}\notag\\ &+\sin{(\omega_{\text{IF}}T(i-1))}\sin{(\omega_{\text{IF}}T(j-1))}\notag\\
&=\cos{\Big(\frac{\pi}{2}(i-1)\Big)}\cos{\Big(\frac{\pi}{2}(j-1)\Big)}\notag\\ &+\sin{\Big(\frac{\pi}{2}(i-1)\Big)}\sin{\Big(\frac{\pi}{2}(j-1)\Big)}.
\end{align}
Fig. \ref{superhet:frequency} visualizes the power spectral density for a superheterodyne front-end and $\kappa=4$, $\text{SNR}=\SI{-3.0}{\decibel}$.

For the considered receiver, the source bandwidth is $B_s\leq\frac{B_y}{2}$, such that $\kappa \geq 2$. For evaluation, we define
\begin{align}\label{def:efficiency}
\chi_i=\frac{ \tilde{\mu}_i(\xi^\ast) }{ {\mu}_i },
\end{align}
to compare the binary latency to the average run-time with high-resolution sampling. Here we assume a binary receiver with $K=\kappa K_0$ such that the absolute observation time $T_o=KT$ of each block stays constant when increasing $\kappa$.
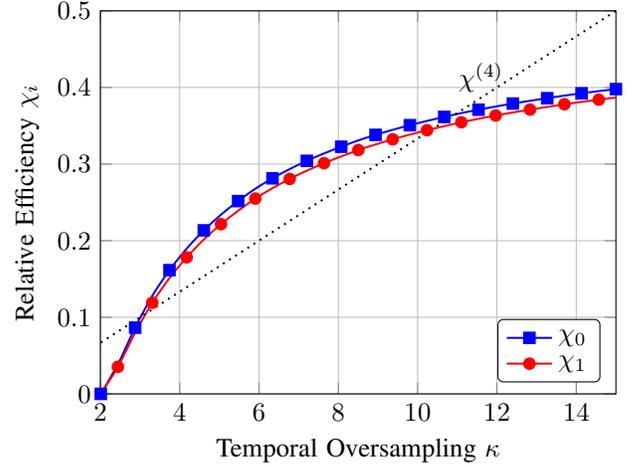
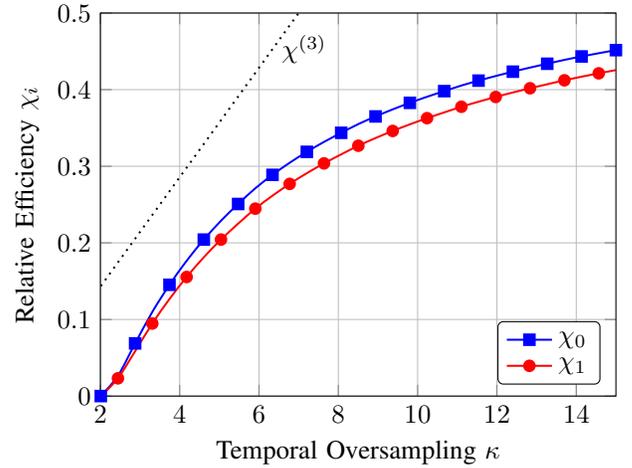
\begin{figure}[!ht]
    \centering
    \begin{subfigure}[t]{0.5\textwidth}
        \centering
        \begin{tikzpicture}[scale=1]

  	\begin{axis}[ylabel=$\text{Relative Efficiency $\chi_i$}$,
  			xlabel=$\text{Temporal Oversampling $\kappa$}$ ,
			grid,
			ymin=0,
			ymax=0.5,
			xmin=2,
			xmax=15,
			legend columns=1,
			legend pos=south east,
			height = 190pt,
			width = 240pt]
						
			\addplot[blue, style=solid, line width=0.75pt, smooth, every mark/.append style={solid}, mark=square*, mark phase = 0, mark repeat=2] table[x index=0, y index=3]{Data/Superhet_Latency_K0_5_SNR_-9_SNRdelta_1.5.txt};
			\addlegendentry{$\chi_0$}
					
			\addplot[red, style=solid, line width=0.75pt, smooth, every mark/.append style={solid}, mark=*, mark phase = 2, mark repeat=2] table[x index=0, y index=4]{Data/Superhet_Latency_K0_5_SNR_-9_SNRdelta_1.5.txt};
			\addlegendentry{$\chi_1$}
			
			\addplot[black, style=dotted, line width=0.75pt, smooth] table[x index=0, y index=3]{Data/efficiency_temporal_superior.txt};
			
			\node[left] at (axis cs: 12.402000000000, 0.413400000000) {$\chi^{(4)}$};
	\end{axis}
	
\end{tikzpicture}
        \caption{Low SNR ($\bar{\theta }=\SI{-9}{\decibel}, \Delta_\theta=\SI{1.5}{\decibel}$)}
        \label{fig:efficiency:bin:cog:low:snr}
    \end{subfigure}
    \begin{subfigure}[t]{0.5\textwidth}
        \centering
        \begin{tikzpicture}[scale=1]

  	\begin{axis}[ylabel=$\text{Relative Efficiency $\chi_i$}$,
  			xlabel=$\text{Temporal Oversampling $\kappa$}$ ,
			grid,
			ymin=0,
			ymax=0.5,
			xmin=2,
			xmax=15,
			legend columns=1,
			legend pos=south east,
			height = 190pt,
			width = 240pt]
						
			\addplot[blue, style=solid, line width=0.75pt, smooth, every mark/.append style={solid}, mark=square*, mark phase = 0, mark repeat=2] table[x index=0, y index=3]{Data/Superhet_Latency_K0_5_SNR_0_SNRdelta_1.5.txt};
			\addlegendentry{$\chi_0$}
					
			\addplot[red, style=solid, line width=0.75pt, smooth, every mark/.append style={solid}, mark=*, mark phase = 2, mark repeat=2] table[x index=0, y index=4]{Data/Superhet_Latency_K0_5_SNR_0_SNRdelta_1.5.txt};
			\addlegendentry{$\chi_1$}
			
			\addplot[black, style=dotted, line width=0.75pt, smooth] table[x index=0, y index=2]{Data/efficiency_temporal_superior.txt};

			\node[right] at (axis cs: 6.340000000000, 0.452857142857) {$\chi^{(3)}$};								
	\end{axis}
	
\end{tikzpicture}
        \caption{Medium SNR ($\bar{\theta }=\SI{0}{\decibel}, \Delta_\theta=\SI{1.5}{\decibel}$)}
        \label{fig:efficiency:bin:cog:medium:snr}
    \end{subfigure}
\caption{Binary Cognitive Radio (Superheterodyne Front-end, $M=1, K_0=5$)}
\end{figure}

Fig.~\ref{fig:efficiency:bin:cog:low:snr} shows \eqref{def:efficiency} for a low SNR scenario with $\bar{\theta }=\SI{-9}{\decibel}$ and $\Delta_\theta=\SI{1.5}{\decibel}$. Without oversampling ($\kappa=2$), binary sampling of the analog sensor signal \eqref{analog:receive:signal:superheterodyne} makes it impossible to perform the detection task. The activity of the transmitter can then only be detected by discriminating between two variance levels. The output of \eqref{system:model:sign}, however, is invariant to changes in its input scale. With oversampling, the presence of a band-limited source introduces correlations among the $K$ samples (non-zero off-diagonal entries in \eqref{covariance:superheterodyne}), which can be detected from the binary measurements. The $\infty$-bit receiver does not benefit from oversampling as the full information about the band-limited sources in \eqref{analog:receive:signal:superheterodyne} is already embedded in the digital measurement data obtained with $\kappa=2$. Fig.~\ref{fig:efficiency:bin:cog:medium:snr} depicts the efficiency \eqref{def:efficiency} at medium SNR ($\bar{\theta }=\SI{0}{\decibel}, \Delta_\theta=\SI{1.5}{\decibel}$). Like for the low SNR case in Fig.~\ref{fig:efficiency:bin:cog:low:snr}, oversampling decreases detection latency. Fig.~\ref{fig:efficiency:bin:cog:low:snr} and Fig.~\ref{fig:efficiency:bin:cog:medium:snr} show that, in general, a binary wireless receiver is suitable for cognitive radio. However, oversampling of the source signal is required, and a larger latency needs to be accepted in return for the simplicity of the radio front-end and the data structure. 

To study complexity, we define the average sampling cost
\begin{align}\label{def:ASC}
\text{ASC}^{(b)}(M,K;\ve{\theta})=\text{SC}^{(b)}(M,K) \cdot \text{ASN}^{(b)}(\ve{\theta}).
\end{align}
A binary receiver outperforms a $b$-bit system in terms of A/D comparator operations when
\begin{align}\label{inequality:superior}
\frac{\text{ASC}^{(1)}(M,K;\ve{\theta})}{\text{ASC}^{(b)}(M,K;\ve{\theta})} < 1.
\end{align}
For a cognitive radio system with the front-end depicted in Fig.~\ref{superhet:receiver}, inequality \eqref{inequality:superior} is fulfilled if
\begin{align}\label{inequality:superior:conservative}
\frac{ \text{SC}^{(1)}(M,\kappa K_0) }{ \text{SC}^{(b)}(M,2K_0)  } \frac{\text{ASN}^{(1)}(\ve{\theta})}{\text{ASN}^{(\infty)}(\ve{\theta})} < 1,
\end{align}
such that a $1$-bit receiver, which exceeds the efficiency level
\begin{align}\label{efficiency:level:superior:superhet}
\chi^{(b)}= \frac{\kappa}{2(2^b-1)}
\end{align}
is superior to a $b$-bit system. Fig.~\ref{fig:efficiency:bin:cog:low:snr} demonstrates that a binary device, in terms of \eqref{def:ASC}, outperforms receivers with $4$ or more bits A/D resolution. Under the benchmark \eqref{inequality:superior:conservative}, Fig.~\ref{fig:efficiency:bin:cog:medium:snr} indicates that a binary system might not be able to compete with a $3$-bit receiver. Note that, concerning the $1$-bit system, \eqref{inequality:superior:conservative} is conservative as we approximate $\text{ASN}^{(b)}\approx\text{ASN}^{(\infty)}$ while $\text{ASN}^{(b)}>\text{ASN}^{(\infty)}$. Also, note that \eqref{inequality:superior:conservative} does not account for the reduced complexity of the analog preprocessing with A/B conversion, e.g., no automatic gain control (AGC) required.
\subsection{High-Performance GNSS Spectrum Monitoring Systems}\label{sec:gnss:monitor}
As a second application, we consider GNSS spectrum monitoring. The task is to detect interference in the vicinity of a satellite radio receiver synchronizing critical infrastructure (e.g., financial market with high-speed trading, supply point of an electrical network) or providing correction data to mobile GNSS receivers which have a strict reliability requirement on their real-time positioning solution (e.g., ground-based augmentation system at an airport). In case the interference on the GNSS band exceeds a certain power level, the monitor issues a warning to the GNSS receiver, which initiates measures to suppress the interference or reports a temporary malfunction.
\begin{figure}[!ht]
    \centering
    \begin{tikzpicture}[scale=1]
	
		% antenna
		\draw (-1.25-0.5,0.75) node[left] {$\breve{y}(t)$};
		\draw [thick,rounded corners=2pt] (-1.0-0.5,0.5) -- (-1.0-0.5,0);
		\draw [thick,rounded corners=2pt] (-1.0-0.5,0.5) -- (-1.25-0.5,0.75);
		\draw [thick,rounded corners=2pt] (-1.0-0.5,0.5) -- (-0.75-0.5,0.75);

		% connection: antenna - LNA
		\draw [thick,rounded corners=2pt,->] (-1.0-0.5,0) -- (-0.5-0.5,0);
		
		% LNA
		\draw (-0.125-0.5,0) node {\footnotesize LNA};
		\draw [thick,rounded corners=2pt] (-0.5-0.5,-0.5) -- (-0.5-0.5,0.5);
		\draw [thick,rounded corners=2pt] (-0.5-0.5,-0.5) -- (0.5-0.5,0);
		\draw [thick,rounded corners=2pt] (-0.5-0.5,0.5) -- (0.5-0.5,0);
		
		% connection: LNA - BP
		\draw [thick,rounded corners=2pt,->] (0.5-0.5,0) -- (1.0-0.5,0);
		
		% BP
		\draw (1.5-0.5,0) node {\footnotesize BP};
		\draw [thick,rounded corners=2pt] (1.0-0.5,-0.4) rectangle (2.0-0.5,0.4);
		
		% connection: BP - LO
		\draw [thick,rounded corners=2pt,->] (2.0-0.5,0) -- (2,0) -- (2,1) -- (2.5,1);
		\draw [thick,rounded corners=2pt,->] (2.0-0.5,0) -- (2,0) -- (2,-1) -- (2.5,-1);
		
		% connection: LO
		\draw (2.75,0+1) node {$\times$};
		\draw (2.75,0+1) circle (0.25cm);
		
		% connection: antenna - LNA
		\draw (2.75,0.5+1) node[above] {\footnotesize $f_{\text{LO}}=f_{\text{C}}$};
		\draw [thick,rounded corners=2pt,->] (2.75,0.5+1) -- (2.75,0.25+1);
		
		% connection:  LO - LP
		\draw [thick,rounded corners=2pt,->] (3.0,0+1) -- (3.5,0+1);
		
		% LP
		\draw (4.0,0+1) node {\footnotesize LP};
		\draw [thick,rounded corners=2pt] (3.5,-0.4+1) rectangle (4.5,0.4+1);
		
		% connection: LP - ADC
		\draw (5.0,0+1) node[right] {${y}_{\text{I}}(t)$};
		\draw [thick,rounded corners=2pt,->] (4.5,0+1) -- (5.0,0+1);
		
		% connection: LO
		\draw (2.75,0-1) node {$\times$};
		\draw (2.75,0-1) circle (0.25cm);
		
		% connection: antenna - LNA
		\draw (2.75,-0.5-1) node[below] {\footnotesize $f_{\text{LO}}=f_{\text{C}}$};
		\draw (2.75,-1.9) node[below] {\footnotesize (with $90^\circ$ phase-shift)};
		\draw [thick,rounded corners=2pt,->] (2.75,-0.5-1) -- (2.75,-0.25-1);
		
		% connection:  LO - LP
		\draw [thick,rounded corners=2pt,->] (3.0,0-1) -- (3.5,0-1);
		
		% LP
		\draw (4.0,0-1) node {\footnotesize LP};
		\draw [thick,rounded corners=2pt] (3.5,-0.4-1) rectangle (4.5,0.4-1);
		
		% connection: LP - ADC
		\draw (5.0,0-1) node[right] {${y}_{\text{Q}}(t)$};
		\draw [thick,rounded corners=2pt,->] (4.5,0-1) -- (5.0,0-1);

	\end{tikzpicture}   
    \caption{Homodyne Analog Sensor Front-End}
    \label{homo:receiver}
\end{figure}
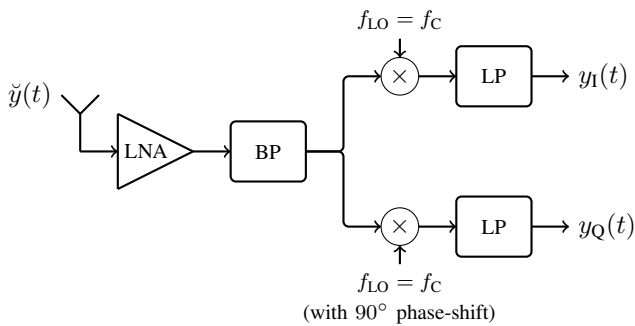

In contrast to the cognitive receiver in Sec.~\ref{sec:cog:radio}, for GNSS monitoring, we assume that the receiver features multiple antennas with a homodyne front-end, as depicted in Fig.~\ref{homo:receiver}. Homodyne front-ends demodulate the received signal within two real-valued channels (in-phase and quadrature) at carrier frequency $f_{\text{C}}$, where the quadrature oscillator features a phase-shift of $90^\circ$ relative to the in-phase demodulator. In the signal processing and communication engineering literature, the two real-valued outputs in Fig.~\ref{homo:receiver} are usually summarized in one complex-valued variable. Note, that this is a mathematical convention that serves compactness. The information embedded in the two signals, and thus the achievable performance does not change when switching between real-valued and complex-valued notation. In fact, complex-valued Gaussian models are usually limited to both variables being uncorrelated and of equivalent variance. Removing these limitations within a complex-valued framework, such that the full information carried by both components can be exploited, requires additional effort concerning notation, see, e.g., \cite{SchreierBook}. A real-valued digital data characterization, which is adjusted to the physical signal acquisition does not face such restrictions and allows analyzing unconventional front-ends, e.g., devices with more than two analog demodulation outputs \cite{Stein15_WCL}. Also, note that with a superheterodyne front-end layout like in Fig.~\ref{superhet:receiver} and $\infty$-bit A/D conversion, one can obtain the digital equivalents of the two analog homodyne outputs in Fig.~\ref{homo:receiver} through digital processing. With low-resolution sampling, however, this is not possible. Consequently, when analyzing homodyne front-ends, we stay in a real-valued notation to keep our framework independent from the specific front-end architecture. So, we denote the analog homodyne outputs \eqref{receive:signal:analog} as $\ve{y}(t)=\begin{bmatrix}\ve{y}^{\T}_\text{I}(t) &\ve{y}^{\T}_\text{Q}(t) \end{bmatrix}^{\T},$ where $\ve{y}_\text{I}(t)\in\fieldR^{M_\text{A}}$ and $\ve{y}_\text{Q}(t)\in\fieldR^{M_\text{A}}$ summarize the analog in-phase and quadrature outputs of the $M_\text{A}=\frac{M}{2}$ receivers. The analog signals have the structure
\begin{align}\label{analog:array:output:structure}
\ve{y}(t)=\theta\ve{A}\ve{x}(t)+\ve{\eta}(t),
\end{align}
where the independent zero-mean random Gaussian processes $\ve{x}(t)=\begin{bmatrix}x_\text{I}(t) &x_\text{Q}(t) \end{bmatrix}^{\T}$ model an interferer with bandwidth $B_s=B_y$ received with an SNR $\theta$. Under the assumption that the bandwidth $B_y$ is narrow in comparison to the carrier frequency $f_{\text{C}}$ and the $M_\text{A}$ sensors are placed as an uniform linear array (ULA) at a distance of half the carrier wavelength, the array steering matrix can be characterized by $\ve{A}=\begin{bmatrix}\ve{A}^{\T}_\text{I} &\ve{A}^{\T}_\text{Q} \end{bmatrix}^{\T}$,
where $\ve{A}_\text{I}, \ve{A}_\text{Q}\in\fieldR^{M_\text{A} \times 2}$ are \cite{Stein16_WSA}
\begin{align}
\ve{A}_\text{I}=\begin{bmatrix}
\cos{\big(0\big)} &\sin{\big(0\big)}\\ 
\cos{\big(\pi\sin{(\varphi)}\big)} &\sin{\big(\pi\sin{(\varphi)}\big)}\\ 
\vdots &\vdots\\ 
\cos{\big((M_\text{A}-1)\pi\sin{(\varphi)}\big)} &\sin{\big((M_\text{A}-1)\pi\sin{(\varphi)}\big)}
\end{bmatrix}
\end{align}
and
\begin{align}
\ve{A}_\text{Q}=\begin{bmatrix}
-\sin{\big(0\big)} &\cos{\big(0\big)}\\ 
-\sin{\big(\pi\sin{(\varphi)}\big)} &\cos{\big(\pi\sin{(\varphi)}\big)} \\ 
\vdots &\vdots\\ 
-\sin{\big((M_\text{A}-1)\pi\sin{(\varphi)}\big)} &\cos{\big((M_\text{A}-1)\pi\sin{(\varphi)}\big)}
\end{bmatrix},
\end{align}
while $\varphi$ denotes the angle under which the source $\ve{x}(t)$ impinges onto the ULA. Assuming that the sampling rate is given by $f_{\text{T}}=\kappa 2B_y$, for a wireless array receiver with an ideal $\infty$-bit A/D conversion process, one obtains zero-mean multivariate Gaussian data \eqref{digital:model:multivariate:gauss} with covariance
\begin{align}\label{covariance:homodyne:array}
\ve{R}_{\ve{y}}(\theta)= \big(\theta \ve{A}\ve{A}^{\T} + \ve{I}\big) \otimes \ve{\Sigma}(\kappa),\quad \ve{R}_{\ve{y}}(\theta)\in\fieldR^{MK \times MK}.
\end{align}
In this case, the noise is not necessarily temporally white as the sampling rate can be misaligned (regarding the sampling theorem) with the analog prefilter bandwidth, i.e., $f_{\text{T}}>2B_y$.

\begin{figure}[!ht]
    \centering
    \begin{tikzpicture}[scale=1]

  	\begin{axis}[ylabel=$\text{Relative Efficiency $\chi_i$}$,
  			xlabel=$\text{Number of Antennas $M_{\text{A}}$}$ ,
			grid,
			ymin=0.1,
			ymax=1.0,
			xmin=2,
			xmax=40,
			legend columns=1,
			legend pos=north west,
			height = 190pt,
			width = 240pt]
						
			\addplot[blue, style=solid, line width=0.75pt, smooth, every mark/.append style={solid}, mark=square*, mark phase = 0, mark repeat=1] table[x index=0, y index=3]{Data/HomodyneArray_Latency_K0_1_SNR_-15_SNRdelta_3_angle_15.txt};
			\addlegendentry{$\chi_0$}
					
			\addplot[red, style=solid, line width=0.75pt, smooth, every mark/.append style={solid}, mark=*, mark phase = 0, mark repeat=1] table[x index=0, y index=4]{Data/HomodyneArray_Latency_K0_1_SNR_-15_SNRdelta_3_angle_15.txt};
			\addlegendentry{$\chi_1$}
			
			\addplot[black, style=dotted, line width=0.75pt, smooth] coordinates {(2,1/3) (40,1/3)};
			
			\node[above] at (axis cs: 37, 0.33) {$\chi^{(2)}$};					
	\end{axis}
	
    \end{tikzpicture}
    \caption{Binary GNSS Monitoring (Homodyne Front-end, $\varphi=15^\circ, \kappa=1, K_0=1, \bar{\theta }=\SI{-15}{\decibel}, \Delta_\theta=\SI{3}{\decibel}$)}
    \label{fig:efficiency:spec:mon:low:snr}
\end{figure}
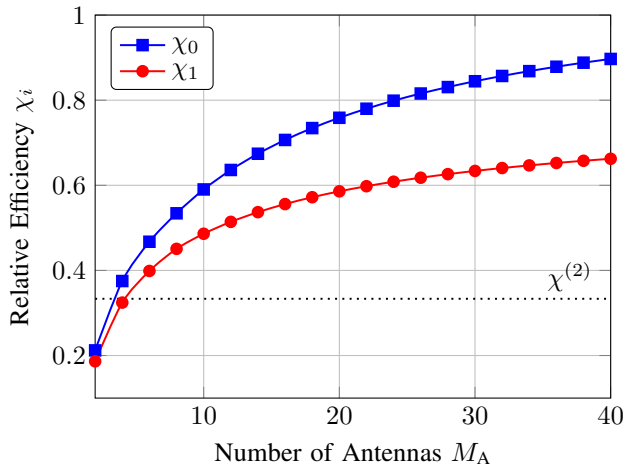

Fig.~\ref{fig:efficiency:spec:mon:low:snr} shows the relative performance measure \eqref{def:efficiency}, where the GNSS radio spectrum is monitored by a binary array with $\bar{\theta }=\SI{-15}{\decibel}$ SNR and $\Delta_\theta=\SI{3}{\decibel}$. No temporal oversampling is performed, i.e., $\kappa=1$. It can be observed that the sensing efficiency of the binary array, in comparison to an ideal $\infty$-bit system, increases with the number of antennas.

\begin{figure}[!ht]
    \centering
    \begin{tikzpicture}[scale=1]

  	\begin{axis}[ylabel=$\text{Relative Efficiency $\chi_{i,m}$}$,
  			xlabel=$\text{Number of Antennas $M_{\text{A}}$}$ ,
			grid,
			ymin=0,
			ymax=1.5,
			xmin=2,
			xmax=40,
			legend columns=2,
			legend pos=south east,
			height = 190pt,
			width = 240pt]
						
			\addplot[blue, style=solid, line width=0.75pt, smooth, every mark/.append style={solid}, mark=square*, mark phase = 0, mark repeat=1] table[x index=0, y index=3]{Data/HomodyneArray_Latency_benchmark_K0_1_SNR_-15_SNRdelta_3_angle_15.txt};
			\addlegendentry{$\chi_{0,4}$}
	
			\addplot[blue, style=solid, line width=0.75pt, smooth, every mark/.append style={solid}, mark=triangle*, mark phase = 0, mark repeat=1] table[x index=0, y index=7]{Data/HomodyneArray_Latency_benchmark_K0_1_SNR_-15_SNRdelta_3_angle_15.txt};
			\addlegendentry{$\chi_{0,16}$}
				
			\addplot[red, style=solid, line width=0.75pt, smooth, every mark/.append style={solid}, mark=*, mark phase = 0, mark repeat=1] table[x index=0, y index=4]{Data/HomodyneArray_Latency_benchmark_K0_1_SNR_-15_SNRdelta_3_angle_15.txt};
			\addlegendentry{$\chi_{1,4}$}
				
			\addplot[red, style=solid, line width=0.75pt, smooth, every mark/.append style={solid}, mark=pentagon*, mark phase = 0, mark repeat=1] table[x index=0, y index=8]{Data/HomodyneArray_Latency_benchmark_K0_1_SNR_-15_SNRdelta_3_angle_15.txt};
			\addlegendentry{$\chi_{1,16}$}
						
			\addplot[black, style=dotted, line width=0.75pt, smooth] table[x index=0, y index=2]{Data/efficiency_spatial_superior.txt};
			
			\addplot[black, style=dotted, line width=0.75pt, smooth] table[x index=0, y index=4]{Data/efficiency_spatial_superior.txt};
			
			\node[below] at (axis cs: 29, 0.604166666667) {$\chi^{(2,16)}$};
			\node[right] at (axis cs: 10, 0.833333333333) {$\chi^{(2,4)}$};	
										
	\end{axis}
	
\end{tikzpicture}
    \caption{Binary GNSS Monitoring (Homodyne Front-end, $\varphi=5^\circ, \kappa=1, K_0=1, \bar{\theta }=\SI{-15}{\decibel}, \Delta_\theta=\SI{3}{\decibel}$)}
    \label{fig:efficiency:spec:mon:low:snr:rel}
\end{figure}
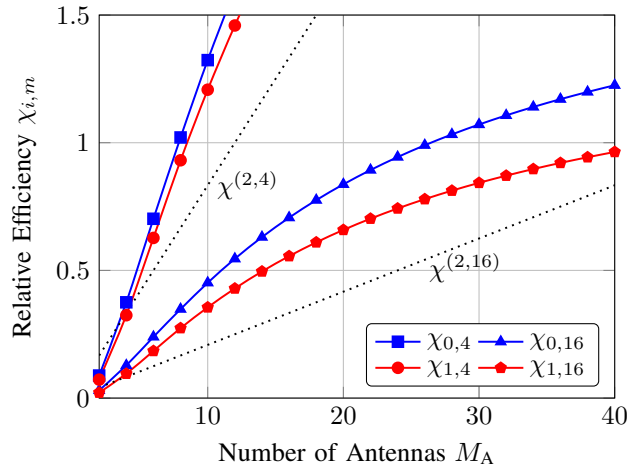

With a homodyne array, the inequality \eqref{inequality:superior} holds when
\begin{align}\label{inequality:superior:conservative:spatial}
\frac{ \text{SC}^{(1)}(M,\kappa K_0) }{ \text{SC}^{(b)}(M,K_0)  } \frac{\text{ASN}^{(1)}(\ve{\theta})}{\text{ASN}^{(\infty)}(\ve{\theta})} < 1.
\end{align}
Therefore, to outperform a $b$-bit homodyne array concerning the A/D cost measure \eqref{def:ASC}, a binary system with the same number of radio antennas needs to exceed the efficiency level
\begin{align}\label{efficiency:level:superior:homodyne}
\chi^{(b)}= \frac{\kappa}{2^b-1}.
\end{align}
The results in Fig.~\ref{fig:efficiency:spec:mon:low:snr} show that binary arrays with more than four homodyne antennas outperform $2$-bit systems regarding digitization complexity when performing spectrum monitoring. To assess how many additional binary sensors are required to outperform an ideal $\infty$-bit array with $m$ radio sensors in terms of latency, we define the relative efficiency
\begin{align}\label{def:efficiency:benchmark}
\chi_{i,m}=\frac{ \tilde{\mu}_i(\xi^\ast) }{ {\mu}_i |_{M_{\text{A}}=m} }.
\end{align}
This measure is depicted in Fig.~\ref{fig:efficiency:spec:mon:low:snr:rel}, where we use two $\infty$-bit ULAs ($m=4$ and $m=16$) as a performance benchmark. A binary array system with $M_{\text{A}}=8$ antennas provides the same sensing latency as a $\infty$-bit A/D resolution system with $m=4$ radio sensors. For outperforming an ideal $\infty$-bit radio system with $m=16$, a binary array with $M_{\text{A}}=40$ sensors is required. Using that \eqref{inequality:superior} holds in the examined scenario if
\begin{align}\label{inequality:superior:conservative:spatial:bench}
\frac{ \text{SC}^{(1)}(2 M_{\text{A}},\kappa K_0) }{ \text{SC}^{(b)}(2 m,K_0)  } \frac{\text{ASN}^{(1)}(\ve{\theta})}{\text{ASN}^{(\infty)}(\ve{\theta})} < 1,
\end{align}
Fig.~\ref{fig:efficiency:spec:mon:low:snr:rel} shows that binary arrays can significantly outperform
\begin{align}\label{efficiency:level:superior:homodyne:benchmark}
\chi^{(b,m)}= \frac{M_{\text{A}}}{m} \frac{\kappa}{(2^b-1)}
\end{align}
evaluated at $b=2$ for the two benchmark systems.

Note that using \eqref{inequality:superior:conservative:spatial} and \eqref{inequality:superior:conservative:spatial:bench} involves underestimating the latency of a $2$-bit system by the ASN of $\infty$-bit receivers such that \eqref{efficiency:level:superior:homodyne} and \eqref{efficiency:level:superior:homodyne:benchmark} form conservative thresholds.
\section{Application - Low-Latency Binary Sensing}
Finally, we investigate the accuracy of our latency analysis via \eqref{kl:01:approx} and \eqref{kl:10:approx} by Monte-Carlo simulations of the binary radio systems considered in Sec.~\ref{sec:application}. To this end, we optimize the ALLR \eqref{lrt:approx} with \eqref{tuning:balance}, run the ASPRT according to \eqref{lrr:exp:approx}, and compare the empirical ASN with the analytic results obtained with \eqref{kl:01:approx} and \eqref{kl:10:approx}. For the simulations, we run each sequential algorithm \num{10000} times with independent observations. The target error rate, determining the thresholds \eqref{sprt:thresh:low} and \eqref{sprt:thresh:up}, is set to $\alpha_0=\alpha_1=0.001$ for all experiments.

\subsection{Low-Latency Decision-Making for Binary Cognitive Radio}
For the cognitive radio setup with superheterodyne front-end  (see Sec. \ref{sec:cog:radio}), the simulation scenario is $M=1, K_0=5, \kappa=5.92$ with the hypotheses at a distance of $\Delta_\theta=\SI{1.5}{\decibel}$ from $\bar{\theta }=\SI{-9}{\decibel}$. Table \ref{simulation:results:superhet:binary} shows a comparison between the analytical and the empirical binary receiver operating characteristics of the ASPRT. For comparison, the $\infty$-bit digitization case, where we employ the exact LLR \eqref{def:llr:exp:fam} and the Gaussian formulas \eqref{kl:divergence:gaussian:h0} and \eqref{kl:divergence:gaussian:h1}, is given in Table~\ref{simulation:results:superhet:ideal}.
\begin{table}[!ht]
    \begin{minipage}{\linewidth}
        \centering
          \begin{tabular}{l c c c c}
    \toprule
    & \multicolumn{4}{c}{\textbf{Receiver Operating Characteristics}} \\ 
    \cmidrule(l){2-5}
    {$\mathcal{H}_i$} & $\alpha_i\text{ (ana.)}$ & $\alpha_i\text{ (sim.)}$ & $\text{ASN}_i\text{ (ana.)}$ & $\text{ASN}_i\text{ (sim.)}$\\ % Column names row
    \midrule
    $\mathcal{H}_0$ & 0.0010 & 0.0010 & 1361.67 & 1367.38\\
    $\mathcal{H}_1$ & 0.0010 & 0.0007 & 1351.21 & 1373.34\\
    \bottomrule\\
  \end{tabular}
        \caption{Binary Cognitive Radio ($\kappa=5.92$)}
        \label{simulation:results:superhet:binary}
    \end{minipage}
    
    \vspace{0.5cm}
    \begin{minipage}{\linewidth}
        \centering
          \begin{tabular}{l c c c c}
    \toprule
    & \multicolumn{4}{c}{\textbf{Receiver Operating Characteristics}} \\
    \cmidrule(l){2-5}
    {$\mathcal{H}_i$} & $\alpha_i\text{ (ana.)}$ & $\alpha_i\text{ (sim.)}$ & $\text{ASN}_i\text{ (ana.)}$ & $\text{ASN}_i\text{ (sim.)}$\\ % Column names row
    \midrule
    $\mathcal{H}_0$ & 0.0010 & 0.0011 & 365.15 & 368.71\\ % Content row 1
    $\mathcal{H}_1$ & 0.0010 & 0.0006 & 344.83 & 351.14\\ % Content row 2
    \bottomrule\\
  \end{tabular}
        \caption{Ideal Cognitive Radio ($\kappa=5.92$)}
        \label{simulation:results:superhet:ideal}
    \end{minipage}
\end{table}
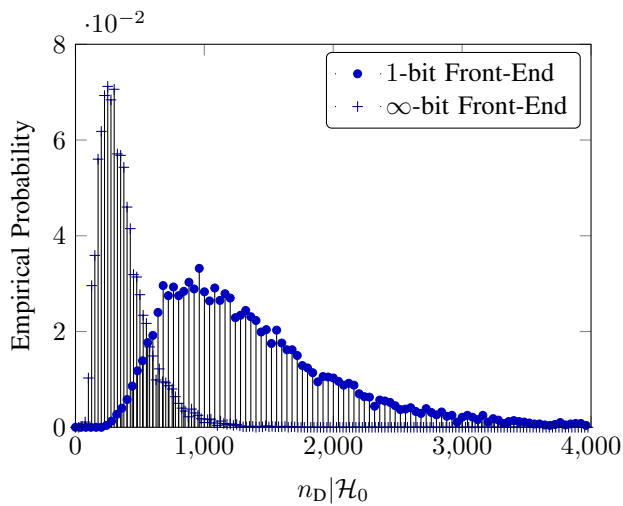
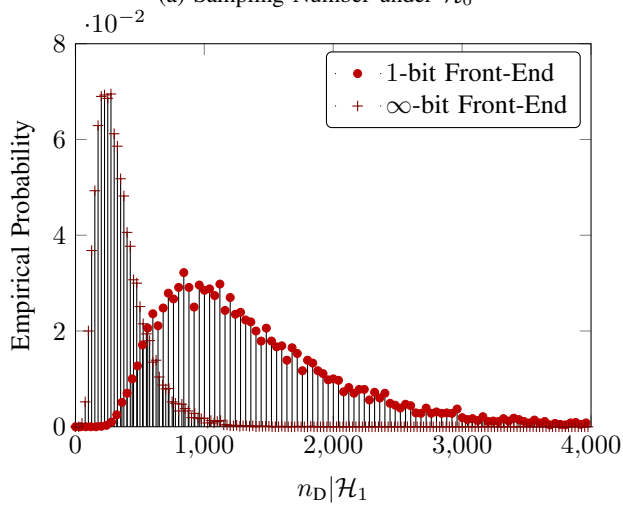
\begin{figure}[!ht]
    \centering
    \begin{subfigure}[t]{0.5\textwidth}
        \centering
          \begin{tikzpicture}[scale=1]
  	\begin{axis}[xlabel=$n_\text{D}|\mathcal{H}_0$,
			ylabel=$\text{Empirical Probability}$,
			ymin=0,
			ymax=0.08,
			xmin=1,
			xmax=4000,
			height = 190pt,
			width = 240pt]

			\addplot+[ycomb, black, mark=*,mark size=1.5pt, mark options={blue!75!black}] table[x index=1, y index=0]{Data/Superhet_Latency_Simulation_K0_5_SNR_-9_SNRdelta_1.5_kappa_5.9161_alpha_0.001_hist.txt};
			\addlegendentry{$1$-bit Front-End}	
	
			\addplot+[ycomb, black, mark=+,mark size=2pt, mark options={blue!50!black}] table[x index=1, y index=0]{Data/Superhet_Latency_Simulation_ideal_K0_5_SNR_-9_SNRdelta_1.5_kappa_5.9161_alpha_0.001_hist.txt};
			\addlegendentry{$\infty$-bit Front-End}
						
	\end{axis}
	
\end{tikzpicture}
        \caption{Sampling Number under $\mathcal{H}_0$}
        \label{fig:asprt:bin:cog:low:snr:H0}
    \end{subfigure}
    \begin{subfigure}[t]{0.5\textwidth}
        \centering
          \begin{tikzpicture}[scale=1]

  	\begin{axis}[xlabel=$n_\text{D}|\mathcal{H}_1$,
			ylabel=$\text{Empirical Probability}$,
			ymin=0,
			ymax=0.08,
			xmin=1,
			xmax=4000,
			height = 190pt,
			width = 240pt]

			\addplot+[ycomb, black, mark=*,mark size=1.5pt, mark options={red!75!black}] table[x index=3, y index=2]{Data/Superhet_Latency_Simulation_K0_5_SNR_-9_SNRdelta_1.5_kappa_5.9161_alpha_0.001_hist.txt};
			\addlegendentry{$1$-bit Front-End}
			
			\addplot+[ycomb, black, mark=+,mark size=2pt, mark options={red!50!black}] table[x index=3, y index=2]{Data/Superhet_Latency_Simulation_ideal_K0_5_SNR_-9_SNRdelta_1.5_kappa_5.9161_alpha_0.001_hist.txt};
			\addlegendentry{$\infty$-bit Front-End}
					
	\end{axis}
	
\end{tikzpicture}
        \caption{Sampling Number under $\mathcal{H}_1$}
        \label{fig:asprt:bin:cog:low:snr:H1}
    \end{subfigure}%
    \caption{Cognitive Radio (Superheterodyne, $M=1, K_0=5, \kappa=5.92, \bar{\theta }=\SI{-9}{\decibel}, \Delta_\theta=\SI{1.5}{\decibel}$)}
\end{figure}

The empirical behavior of both tests shows a good correspondence with the analytic results. For further illustration, the empirical distribution $n_{\text{D}}$ under both hypotheses and digitization approaches is depicted in Fig.~\ref{fig:asprt:bin:cog:low:snr:H0} and Fig.~\ref{fig:asprt:bin:cog:low:snr:H1}. The results corroborate that a binary cognitive receiver can reliably sense the activity of a weak primary user through temporal oversampling. As predicted by the analysis in Fig.~\ref{fig:efficiency:bin:cog:low:snr}, the binary system requires significantly more observation time than the ideal receiver to obtain the specified reliability level.
\subsection{Low-Latency Detection for Binary GNSS Monitoring}
For the second set of simulations, we consider a binary ULA with  $M_{\text{A}}=8$ homodyne front-ends as analyzed in Sec.~\ref{sec:gnss:monitor}. The direction-of-arrival is $\varphi=5^\circ$ and the sampling configuration $\kappa=1, K_0=1$. The interference detection is performed by centering the two hypotheses around $\bar{\theta }=\SI{-15}{\decibel}$ at a distance of  $\Delta_\theta=\SI{3}{\decibel}$. For comparison, an $\infty$-bit array system with half the amount of wireless sensors ($M_{\text{A}}=4$) is used. Table~\ref{simulation:results:homo:binary} and Table~\ref{simulation:results:homo:ideal} show the obtained results where the empirical outcomes match the analytical assessments.
\begin{table}[!ht]
    \begin{minipage}{\linewidth}
        \centering
          \begin{tabular}{l c c c c}
    \toprule
    & \multicolumn{4}{c}{\textbf{Receiver Operating Characteristics}} \\
    \cmidrule(l){2-5}
    {$\mathcal{H}_i$} & $\alpha_i\text{ (ana.)}$ & $\alpha_i\text{ (sim.)}$ & $\text{ASN}_i\text{ (ana.)}$ & $\text{ASN}_i\text{ (sim.)}$\\
    \midrule
    $\mathcal{H}_0$ & 0.0010 & 0.0002 & 179.31 & 183.48\\
    $\mathcal{H}_1$ & 0.0010 & 0.0010 & 162.20 & 168.09\\
    \bottomrule\\
  \end{tabular}
        \caption{Binary GNSS Monitoring ($M_{\text{A}}=8$)}
        \label{simulation:results:homo:binary}
    \end{minipage}

    \vspace{0.5cm}
    \begin{minipage}{\linewidth}
        \centering
          \begin{tabular}{l c c c c}
    \toprule
    & \multicolumn{4}{c}{\textbf{Receiver Operating Characteristics}} \\
    \cmidrule(l){2-5}
    {$\mathcal{H}_i$} & $\alpha_i\text{ (ana.)}$ & $\alpha_i\text{ (sim.)}$ & $\text{ASN}_i\text{ (ana.)}$ & $\text{ASN}_i\text{ (sim.)}$\\
    \midrule
    $\mathcal{H}_0$ & 0.0010 & 0.0007 & 182.92 & 184.48\\
    $\mathcal{H}_1$ & 0.0010 & 0.0014 & 150.99 & 159.43\\
    \bottomrule\\
  \end{tabular}
        \caption{Ideal GNSS Monitoring ($M_{\text{A}}=4$)}
        \label{simulation:results:homo:ideal}
    \end{minipage}
\end{table}

\begin{figure}[!ht]
    \centering
    \begin{subfigure}[t]{0.5\textwidth}
        \centering
        \begin{tikzpicture}[scale=1]
  	    \begin{axis}[xlabel=$n_\text{D}|\mathcal{H}_0$,
			ylabel=$\text{Empirical Probability}$,
			ymin=0,
			ymax=0.09,
			xmin=1,
			xmax=800,
			height = 190pt,
			width = 240pt]

			\addplot+[ycomb, black, mark=*,mark size=1.5pt, mark options={blue!75!black}] table[x index=1, y index=0]{Data/HomodyneArray_Latency_Simulation_K0_1_ant_8_SNR_-15_SNRdelta_3_kappa_1_alpha_0.001_hist.txt};
			\addlegendentry{$1$-bit Front-End ($M_{\text{A}}=8$)}	
	
			\addplot+[ycomb, black, mark=+,mark size=2pt, mark options={blue!50!black}] table[x index=1, y index=0]{Data/HomodyneArray_Latency_Simulation_ideal_K0_1_ant_4_SNR_-15_SNRdelta_3_kappa_1_alpha_0.001_hist.txt};
			\addlegendentry{$\infty$-bit Front-End ($M_{\text{A}}=4$)}
						
	    \end{axis}
        \end{tikzpicture}
        \caption{Sampling Number under $\mathcal{H}_0$}
        \label{fig:asprt:bin:gnss:low:snr:H0}
        \end{subfigure}
        \begin{subfigure}[t]{0.5\textwidth}
        \centering
        \begin{tikzpicture}[scale=1]

  	    \begin{axis}[xlabel=$n_\text{D}|\mathcal{H}_1$,
			ylabel=$\text{Empirical Probability}$,
			ymin=0,
			ymax=0.09,
			xmin=1,
			xmax=800,
			height = 190pt,
			width = 240pt]

			\addplot+[ycomb, black, mark=*,mark size=1.5pt, mark options={red!75!black}] table[x index=3, y index=2]{Data/HomodyneArray_Latency_Simulation_K0_1_ant_8_SNR_-15_SNRdelta_3_kappa_1_alpha_0.001_hist.txt};
			\addlegendentry{$1$-bit Front-End ($M_{\text{A}}=8$)}	
			
			\addplot+[ycomb, black, mark=+,mark size=2pt, mark options={red!50!black}] table[x index=3, y index=2]{Data/HomodyneArray_Latency_Simulation_ideal_K0_1_ant_4_SNR_-15_SNRdelta_3_kappa_1_alpha_0.001_hist.txt};
			\addlegendentry{$\infty$-bit Front-End ($M_{\text{A}}=4$)}
					
	    \end{axis}
	
        \end{tikzpicture}
        \caption{Sampling Number under $\mathcal{H}_1$}
        \label{fig:asprt:bin:gnss:low:snr:H1}
        \end{subfigure}%
        \caption{GNSS Monitoring (Homodyne, $\varphi=5^\circ, \kappa=1, K_0=1, \bar{\theta }=\SI{-15}{\decibel}, \Delta_\theta=\SI{3}{\decibel}$)}
\end{figure}
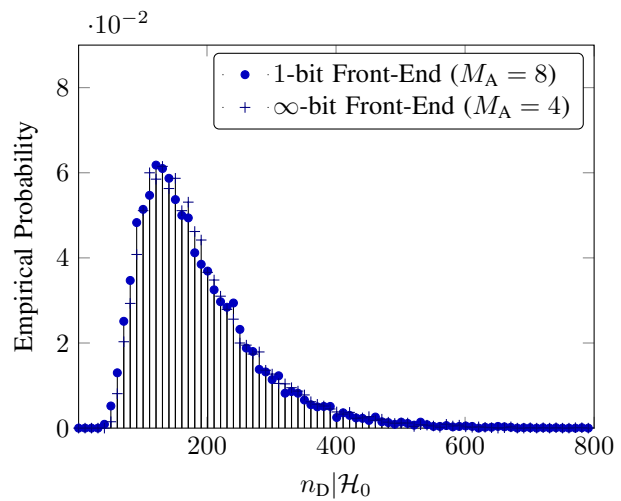
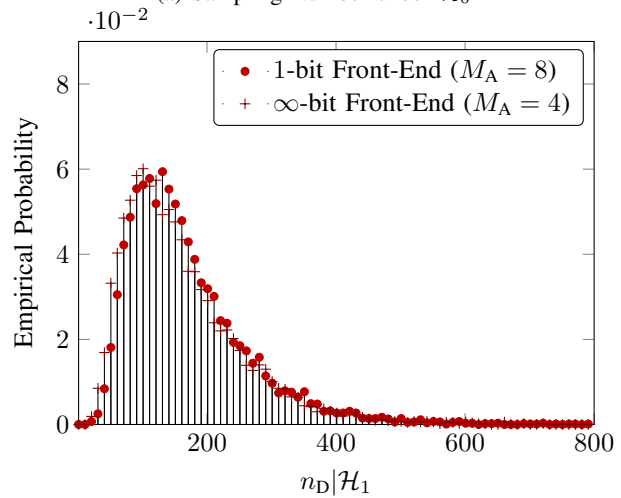

Fig.~\ref{fig:asprt:bin:gnss:low:snr:H0} and Fig.~\ref{fig:asprt:bin:gnss:low:snr:H1} show the empirical distribution of $n_{\text{D}}$. These results confirm that double the amount of binary sensors can be sufficient to obtain a sensing apparatus equivalent to an ideal system with $\infty$-bit A/D resolution. Note that \eqref{efficiency:level:superior:homodyne:benchmark} shows that the considered binary array ($M_{\text{A}}=8$) provides this excellent performance at less than $\SI{66}{\percent}$ of the digitization cost than any other system with $M_{\text{A}}=4$ and a higher A/D resolution. Nevertheless, it requires more space, which is not reflected in our definition of digitization complexity \eqref{def:sample:cost}.
\section{Conclusion}
We have discussed sequential detection with measurements from binary radio systems. Considering statistical tests in the exponential family, we circumvent the intractable distribution model arising when hard-limiting multivariate Gaussian data. Under this general probabilistic framework, we have derived approximations for the log-likelihood ratio and the Kullback--Leibler divergence. The expressions have the advantage that they cover a broad range of different sensor data models and can be adapted to the testing scenario. This provides stronger approximations than fixed methods. For demonstrating the practical impact, we have applied these theoretical tools to the system design specification of wireless systems with binary sensor signal digitization. The results show that radio systems with low-complexity front-ends are capable of performing challenging sequential detection tasks. In particular, using a large number of binary sensors shows to be a favorable approach concerning detection latency and digitization resources. Finally, we have demonstrated that our analysis matches the empirical behavior of sequential detectors operating based on binary radio data streams. In summary, our discussion provides a versatile framework for sequential tests with hardware-aware probabilistic modeling of the sensor data streams and the rule of thumb that for reaching a favorable complexity-latency trade-off in modern sensing architectures, operating in noisy environments, one should reduce the A/D amplitude resolution to a single bit and double the number of sensor devices.
\appendix
\section{Finite Difference Approximations}
For $f: \fieldR \to \fieldR$, infinitely differentiable at $\hat{u}\in\fieldR$, with $\Delta\in\fieldR, \Delta \geq 0,$ by the Taylor series
\begin{align}
f(\hat{u}+\Delta)&=\sum_{i=0}^{\infty} \frac{\Delta^i}{i!} \frac{\partial^i f(\hat{u})}{\partial u^i},\\
f(\hat{u}-\Delta)&=\sum_{i=0}^{\infty} (-1)^i \frac{\Delta^i}{i!} \frac{\partial^i f(\hat{u})}{\partial u^i}.
\end{align}
A forward finite difference approximation at $\hat{u}$ results in
\begin{align}\label{dif:approx:forward}
f(\hat{u}+\Delta)-f(\hat{u}) \approx \frac{\partial f(\hat{u})}{\partial u }  \Delta.
\end{align}
With a backward finite difference approximation at $\hat{u}$
\begin{align}\label{dif:approx:backward}
f(\hat{u})-f(\hat{u}-\Delta) \approx \frac{\partial f(\hat{u})}{\partial u }  \Delta.
\end{align}
Defining $\Delta_a, \Delta_b\in\fieldR$ where $\Delta_a, \Delta_b \geq 0$, an alternative approximation at $\hat{u}$ is
\begin{align}\label{dif:approx:flex}
f(\hat{u}+\Delta_a)-f(\hat{u}-\Delta_b) \approx \frac{\partial f(\hat{u})}{\partial u } (\Delta_a+\Delta_b).
\end{align}
Defining $u_1\geq u_0$, with the forward approximation \eqref{dif:approx:forward}
\begin{align}\label{dif:approx:forward2}
f(u_1)-f(u_0) \approx \frac{\partial f(u_0)}{\partial u }  (u_1-u_0),
\end{align}
and with the backward approximation \eqref{dif:approx:backward}
\begin{align}\label{dif:approx:backward2}
f(u_1)-f(u_0) \approx \frac{\partial f(u_1)}{\partial u }  (u_1-u_0).
\end{align}
Defining $\xi\in [0; 1]$ and
\begin{align}
\hat{u}&=\xi u_0 + (1-\xi) u_1,\\
\Delta_a&=\xi (u_1-u_0),\\
\Delta_b&=(1-\xi) (u_1-u_0),
\end{align}
with \eqref{dif:approx:flex}
\begin{align}\label{dif:approx:flex2}
f(u_1)-f(u_0) \approx \frac{\partial f(\xi u_0 + (1-\xi) u_1)}{\partial u } (u_1-u_0).
\end{align}
Extending \eqref{dif:approx:forward2} - \eqref{dif:approx:flex2} to $\ve{f}: \fieldR^I \to \fieldR^J$ with $I,J\in\fieldN$
\begin{align}\label{dif:approx:forward:vec}
\ve{f}(\ve{u}_1)-\ve{f}(\ve{u}_0) &\approx \frac{\partial \ve{f}(\ve{u}_0)}{\partial \ve{u} }  (\ve{u}_1-\ve{u}_0),\\
\label{dif:approx:backward:vec}
\ve{f}(\ve{u}_1)-\ve{f}(\ve{u}_0) &\approx \frac{\partial \ve{f}(\ve{u}_1)}{\partial \ve{u} }  (\ve{u}_1-\ve{u}_0),\\
\label{dif:approx:flex:vec}
\ve{f}(\ve{u}_1)-\ve{f}(\ve{u}_0) &\approx \frac{\partial \ve{f}(\xi \ve{u}_0 + (1-\xi) \ve{u}_1)}{\partial \ve{u} } (\ve{u}_1-\ve{u}_0).
\end{align}

\balance
\bibliographystyle{IEEEtran}
\bibliography{IEEEabrv,Bibliography}

\begin{thebibliography}{10}
\providecommand{\url}[1]{#1}
\csname url@samestyle\endcsname
\providecommand{\newblock}{\relax}
\providecommand{\bibinfo}[2]{#2}
\providecommand{\BIBentrySTDinterwordspacing}{\spaceskip=0pt\relax}
\providecommand{\BIBentryALTinterwordstretchfactor}{4}
\providecommand{\BIBentryALTinterwordspacing}{\spaceskip=\fontdimen2\font plus
\BIBentryALTinterwordstretchfactor\fontdimen3\font minus
  \fontdimen4\font\relax}
\providecommand{\BIBforeignlanguage}[2]{{%
\expandafter\ifx\csname l@#1\endcsname\relax
\typeout{** WARNING: IEEEtran.bst: No hyphenation pattern has been}%
\typeout{** loaded for the language `#1'. Using the pattern for}%
\typeout{** the default language instead.}%
\else
\language=\csname l@#1\endcsname
\fi
#2}}
\providecommand{\BIBdecl}{\relax}
\BIBdecl

\bibitem{MurmannSurvey}
\BIBentryALTinterwordspacing
B.~{Murmann}. {ADC} performance survey 1997-2018. [Online]. Available:
  \url{http://web.stanford.edu/\%7emurmann/adcsurvey.html}
\BIBentrySTDinterwordspacing

\bibitem{SteinICASSP17}
M.~S. {Stein}, ``Performance analysis for time-of-arrival estimation with
  oversampled low-complexity 1-bit {A/D} conversion,'' in \emph{IEEE Int.
  Conference on Acoustics, Speech and Signal Processing (ICASSP)}, Mar. 2017,
  pp. 4491--4495.

\bibitem{Ivrlac06}
M.~T. {Ivrlac} and J.~A. {Nossek}, ``Challenges in coding for quantized {MIMO}
  systems,'' in \emph{IEEE Int. Symposium on Information Theory}, July 2006,
  pp. 2114--2118.

\bibitem{Dabeer06}
O.~{Dabeer}, J.~{Singh}, and U.~{Madhow}, ``On the limits of communication
  performance with one-bit analog-to-digital conversion,'' in \emph{IEEE
  Workshop on Signal Processing Advances in Wireless Communications}, July
  2006, pp. 1--5.

\bibitem{Mezghani07}
A.~{Mezghani} and J.~A. {Nossek}, ``On ultra-wideband {MIMO} systems with 1-bit
  quantized outputs: Performance analysis and input optimization,'' in
  \emph{IEEE Int. Symposium on Information Theory}, June 2007, pp. 1286--1289.

\bibitem{Mezghani12}
\BIBentryALTinterwordspacing
------, ``Capacity lower bound of {MIMO} channels with output quantization and
  correlated noise,'' July 2012, presented at {IEEE} Int. Symposium on
  Information Theory. [Online]. Available:
  \url{http://mediatum.ub.tum.de/1171263}
\BIBentrySTDinterwordspacing

\bibitem{Mo15}
J.~{Mo} and R.~W. {Heath}, ``Capacity analysis of one-bit quantized {MIMO}
  systems with transmitter channel state information,'' \emph{{IEEE} Trans.
  Signal Process.}, vol.~63, no.~20, pp. 5498--5512, Oct. 2015.

\bibitem{Jacobsson15}
S.~{Jacobsson}, G.~{Durisi}, M.~{Coldrey}, U.~{Gustavsson}, and C.~{Studer},
  ``One-bit massive {MIMO}: Channel estimation and high-order modulations,'' in
  \emph{IEEE Int. Conference on Communication Workshop (ICCW)}, June 2015, pp.
  1304--1309.

\bibitem{Mollen17}
C.~{Moll\'{e}n}, J.~{Choi}, E.~G. {Larsson}, and R.~W. {Heath}, ``Uplink
  performance of wideband massive {MIMO} with one-bit {ADCs},'' \emph{{IEEE}
  Trans. Wireless Commun.}, vol.~16, no.~1, pp. 87--100, Jan. 2017.

\bibitem{Jeon18}
Y.~{Jeon}, N.~{Lee}, S.~{Hong}, and R.~W. {Heath}, ``One-bit sphere decoding
  for uplink massive {MIMO} systems with one-bit adcs,'' \emph{{IEEE} Trans.
  Wireless Commun.}, vol.~17, no.~7, pp. 4509--4521, July 2018.

\bibitem{Daubechies03}
I.~Daubechies and R.~DeVore, ``Approximating a bandlimited function using very
  coarsely quantized data: A family of stable sigma-delta modulators of
  arbitrary order,'' \emph{Ann. Math.}, vol. 158, no.~2, pp. 679--710, 2003.

\bibitem{Kamilov12}
U.~S. {Kamilov}, A.~{Bourquard}, A.~{Amini}, and M.~{Unser}, ``One-bit
  measurements with adaptive thresholds,'' \emph{{IEEE} Signal Process. Lett.},
  vol.~19, no.~10, pp. 607--610, Oct. 2012.

\bibitem{Boufounos15}
P.~T. Boufounos, L.~Jacques, F.~Krahmer, and R.~Saab, \emph{Quantization and
  Compressive Sensing}.\hskip 1em plus 0.5em minus 0.4em\relax Cham: Springer
  International Publishing, 2015, pp. 193--237.

\bibitem{Madsen00}
A.~{Host-Madsen} and P.~{Handel}, ``Effects of sampling and quantization on
  single-tone frequency estimation,'' \emph{{IEEE} Trans. Signal Process.},
  vol.~48, no.~3, pp. 650--662, Mar. 2000.

\bibitem{Ribeiro06_part1}
A.~{Ribeiro} and G.~B. {Giannakis}, ``Bandwidth-constrained distributed
  estimation for wireless sensor networks - {Part I}: Gaussian case,''
  \emph{{IEEE} Trans. Signal Process.}, vol.~54, no.~3, pp. 1131--1143, Mar.
  2006.

\bibitem{Mezghani10}
A.~{Mezghani}, F.~{Antreich}, and J.~A. {Nossek}, ``Multiple parameter
  estimation with quantized channel output,'' in \emph{Int. ITG Workshop on
  Smart Antennas (WSA)}, Feb. 2010, pp. 143--150.

\bibitem{Stein15_WCL}
M.~{Stein}, S.~{Theiler}, and J.~A. {Nossek}, ``Overdemodulation for
  high-performance receivers with low-resolution {ADC},'' \emph{{IEEE} Wireless
  Commun. Lett.}, vol.~4, no.~2, pp. 169--172, Apr. 2015.

\bibitem{Li17}
Y.~{Li}, C.~{Tao}, G.~{Seco-Granados}, A.~{Mezghani}, A.~L. {Swindlehurst}, and
  L.~{Liu}, ``Channel estimation and performance analysis of one-bit massive
  {MIMO} systems,'' \emph{{IEEE} Trans. Signal Process.}, vol.~65, no.~15, pp.
  4075--4089, Aug. 2017.

\bibitem{Mezghani18}
A.~{Mezghani} and A.~L. {Swindlehurst}, ``Blind estimation of sparse broadband
  massive {MIMO} channels with ideal and one-bit {ADCs},'' \emph{{IEEE} Trans.
  Signal Process.}, vol.~66, no.~11, pp. 2972--2983, June 2018.

\bibitem{Willett95}
P.~{Willett} and P.~F. {Swaszek}, ``On the performance degradation from one-bit
  quantized detection,'' \emph{{IEEE} Trans. Inf. Theory}, vol.~41, no.~6, pp.
  1997--2003, Nov. 1995.

\bibitem{Ciuonzo13}
D.~{Ciuonzo}, G.~{Papa}, G.~{Romano}, P.~{Salvo Rossi}, and P.~{Willett},
  ``One-bit decentralized detection with a {Rao} test for multisensor fusion,''
  \emph{{IEEE} Signal Process. Lett.}, vol.~20, no.~9, pp. 861--864, Sep. 2013.

\bibitem{Stein18_ICASSP}
M.~S. {Stein}, ``Asymptotic signal detection rates with 1-bit array
  measurements,'' in \emph{IEEE Int. Conference on Acoustics, Speech and Signal
  Processing (ICASSP)}, Apr. 2018, pp. 4534--4538.

\bibitem{Tantaratana77}
S.~{Tantaratana} and J.~{Thomas}, ``Quantization for sequential signal
  detection,'' \emph{{IEEE} Trans. Commun.}, vol.~25, no.~7, pp. 696--703, July
  1977.

\bibitem{Wald45}
A.~Wald, ``Sequential tests of statistical hypotheses,'' \emph{Ann. Math.
  Statist.}, vol.~16, no.~2, pp. 117--186, 1945.

\bibitem{Fauss15}
M.~Fau{\ss} and A.~M. Zoubir, ``A linear programming approach to sequential
  hypothesis testing,'' \emph{Sequential Analysis}, vol.~34, no.~2, pp.
  235--263, 2015.

\bibitem{Axell12}
E.~{Axell}, G.~{Leus}, E.~G. {Larsson}, and H.~V. {Poor}, ``Spectrum sensing
  for cognitive radio: State-of-the-art and recent advances,'' \emph{{IEEE}
  Signal Process. Mag.}, vol.~29, no.~3, pp. 101--116, May 2012.

\bibitem{Broumandan16}
A.~{Broumandan}, A.~{Jafarnia-Jahromi}, S.~{Daneshmand}, and G.~{Lachapelle},
  ``Overview of spatial processing approaches for {GNSS} structural
  interference detection and mitigation,'' \emph{Proc. {IEEE}}, vol. 104,
  no.~6, pp. 1246--1257, June 2016.

\bibitem{Ioannides16}
R.~T. {Ioannides}, T.~{Pany}, and G.~{Gibbons}, ``Known vulnerabilities of
  global navigation satellite systems, status, and potential mitigation
  techniques,'' \emph{Proc. {IEEE}}, vol. 104, no.~6, pp. 1174--1194, June
  2016.

\bibitem{Hashemi89}
H.~R. {Hashemi} and I.~B. {Rhodes}, ``Decentralized sequential detection,''
  \emph{{IEEE} Trans. Inf. Theory}, vol.~35, no.~3, pp. 509--520, May 1989.

\bibitem{Veeravalli93}
V.~V. {Veeravalli}, T.~{Basar}, and H.~V. {Poor}, ``Decentralized sequential
  detection with a fusion center performing the sequential test,'' \emph{{IEEE}
  Trans. Inf. Theory}, vol.~39, no.~2, pp. 433--442, Mar. 1993.

\bibitem{Hussain94}
A.~M. {Hussain}, ``Multisensor distributed sequential detection,'' \emph{{IEEE}
  Trans. Aerosp. Electron. Syst.}, vol.~30, no.~3, pp. 698--708, July 1994.

\bibitem{Mei08}
Y.~{Mei}, ``Asymptotic optimality theory for decentralized sequential
  hypothesis testing in sensor networks,'' \emph{{IEEE} Trans. Inf. Theory},
  vol.~54, no.~5, pp. 2072--2089, May 2008.

\bibitem{Yilmaz12}
Y.~{Yilmaz}, G.~V. {Moustakides}, and X.~{Wang}, ``Cooperative sequential
  spectrum sensing based on level-triggered sampling,'' \emph{{IEEE} Trans.
  Signal Process.}, vol.~60, no.~9, pp. 4509--4524, Sep. 2012.

\bibitem{Chaudhari12}
S.~{Chaudhari}, J.~{Lundén}, and V.~{Koivunen}, ``Effects of quantization and
  channel errors on sequential detection in cognitive radios,'' in \emph{Annual
  Conference on Information Sciences and Systems (CISS)}, Mar. 2012, pp. 1--6.

\bibitem{Wang13}
Y.~{Wang} and Y.~{Mei}, ``Quantization effect on the log-likelihood ratio and
  its application to decentralized sequential detection,'' \emph{{IEEE} Trans.
  Signal Process.}, vol.~61, no.~6, pp. 1536--1543, Mar. 2013.

\bibitem{Blum95}
R.~S. {Blum}, ``Quantization in multisensor random signal detection,''
  \emph{{IEEE} Trans. Inf. Theory}, vol.~41, no.~1, pp. 204--215, Jan. 1995.

\bibitem{Nguyen08}
X.~{Nguyen}, M.~J. {Wainwright}, and M.~I. {Jordan}, ``On optimal quantization
  rules for some problems in sequential decentralized detection,'' \emph{{IEEE}
  Trans. Inf. Theory}, vol.~54, no.~7, pp. 3285--3295, July 2008.

\bibitem{Teng13}
D.~{Teng} and E.~{Ertin}, ``Optimal quantization of likelihood for low
  complexity sequential testing,'' in \emph{IEEE Global Conference on Signal
  and Information Processing}, Dec. 2013, pp. 675--678.

\bibitem{Tantaratana77_TIT}
S.~{Tantaratana} and J.~{Thomas}, ``On sequential sign detection of a constant
  signal,'' \emph{{IEEE} Trans. Inf. Theory}, vol.~23, no.~3, pp. 304--315, May
  1977.

\bibitem{Lee81}
C.~C. {Lee} and J.~B. {Thomas}, ``Sequential detection based on simple
  quantization,'' \emph{J. Franklin Inst.}, vol. 312, pp. 119--135, 1981.

\bibitem{Choi16}
J.~{Choi}, J.~{Mo}, and R.~W. {Heath}, ``Near maximum-likelihood detector and
  channel estimator for uplink multiuser massive {MIMO} systems with one-bit
  {ADC}s,'' \emph{{IEEE} Trans. Commun.}, vol.~64, no.~5, pp. 2005--2018, May
  2016.

\bibitem{Hong18}
S.~{Hong} and N.~{Lee}, ``Soft-output detector for uplink {MU-MIMO} systems
  with one-bit {ADCs},'' \emph{{IEEE} Commun. Lett.}, vol.~22, no.~5, pp.
  930--933, May 2018.

\bibitem{SteinBar18}
M.~S. Stein, S.~Bar, J.~A. Nossek, and J.~Tabrikian, ``Performance analysis for
  channel estimation with 1-bit {ADC} and unknown quantization threshold,''
  \emph{{IEEE} Trans. Signal Process.}, vol.~66, no.~10, pp. 2557--2571, May
  2018.

\bibitem{Dai13}
B.~Dai, S.~Ding, and G.~Wahba, ``Multivariate {Bernoulli} distribution,''
  \emph{Bernoulli}, vol.~19, no.~4, pp. 1465--1483, 2013.

\bibitem{Gupta63}
S.~S. Gupta, ``Probability integrals of multivariate normal and multivariate
  $t^1$,'' \emph{Ann. Math. Statist.}, vol.~34, no.~3, pp. 792--828, 09 1963.

\bibitem{Stein_FisherBound}
\BIBentryALTinterwordspacing
M.~S. Stein, J.~A. Nossek, and K.~Barb{\'{e}}, ``Fisher information lower
  bounds with applications in hardware-aware nonlinear signal processing,''
  2015. [Online]. Available: \url{http://arxiv.org/abs/1512.03473v2}
\BIBentrySTDinterwordspacing

\bibitem{SteinFauss18}
M.~S. {Stein} and M.~{Fau{\ss}}, ``In a one-bit rush: Low-latency wireless
  spectrum monitoring with binary sensor arrays,'' in \emph{IEEE Statistical
  Signal Processing Workshop (SSP)}, June 2018, pp. 223--227.

\bibitem{TartakovskyBook}
A.~Tartakovsky, I.~Nikiforov, and M.~Basseville, \emph{Sequential Analysis:
  Hypothesis Testing and Changepoint Detection}.\hskip 1em plus 0.5em minus
  0.4em\relax Chapman \& Hall, 2014.

\bibitem{Cox94}
D.~R. Cox and N.~Wermuth, ``A note on the quadratic exponential binary
  distribution,'' \emph{Biometrika}, vol.~81, no.~2, pp. 403--408, 1994.

\bibitem{Stein16_WSA}
M.~{Stein}, K.~{Barb\'{e}}, and J.~A. {Nossek}, ``{DOA} parameter estimation
  with 1-bit quantization - {B}ounds, methods and the exponential
  replacement,'' in \emph{Int. ITG Workshop on Smart Antennas (WSA)}, Mar.
  2016, pp. 1--6.

\bibitem{ThomasBook}
J.~B. {Thomas}, \emph{An introduction to statistical communication
  theory}.\hskip 1em plus 0.5em minus 0.4em\relax Wiley, 1969.

\bibitem{Sinn11}
M.~Sinn and K.~Keller, ``Covariances of zero crossings in {Gaussian}
  processes,'' \emph{Theory of Probability \& Its Applications}, vol.~55,
  no.~3, pp. 485--504, 2011.

\bibitem{Khintchine34}
A.~Khintchine, ``Korrelationstheorie der station{\"a}ren stochastischen
  {P}rozesse,'' \emph{Mathematische Annalen}, vol. 109, no.~1, pp. 604--615,
  Dec. 1934.

\bibitem{SchreierBook}
P.~J. {Schreier} and L.~L. {Scharf}, \emph{Statistical Signal Processing of
  Complex-Valued Data: The Theory of Improper and Noncircular Signals}.\hskip
  1em plus 0.5em minus 0.4em\relax Cambridge University Press, 2010.

\end{thebibliography}
\end{document}